\documentclass[twocolumn]{aastex7}

\usepackage{graphicx}	
\usepackage{amsmath}	
\usepackage{amssymb}	

\newcommand\Mdot{\dot{M}}
\newcommand\MdotB{\dot{M}_\mathrm{B}}
\newcommand\mdot{\dot{m}}

\newcommand\mdoto{\dot{m}_\mathrm{0}}
\newcommand\mdotBH{\dot{m}_\mathrm{BH}}
\newcommand\rB{r_\mathrm{B}}
\newcommand\rcr{r_\mathrm{cr}}
\newcommand\rout{r_\mathrm{out}}
\newcommand\rs{r_\mathrm{s}}
\newcommand\cs{c_\mathrm{s}}
\newcommand\JB{J_\mathrm{B}}
\newcommand\JO{J_\mathrm{0}}
\newcommand\jcr{j_\mathrm{cr}}
\newcommand\Jout{J_\mathrm{out}}
\newcommand\jout{j_\mathrm{out}}
\newcommand\lcr{\lambda_\mathrm{cr}}
\newcommand\rsc{r_\mathrm{s}c}

\newcommand\lBB{\lambda_\mathrm{{\tiny BB}}}

\begin{document}

\title{Generalized Bondi Accretion Flow with and without Outflow}

\correspondingauthor{Myeong-Gu Park}
\email{mgp@knu.ac.kr}

\author[orcid=0000-0003-3939-8610]{Dongho Han}
\affiliation{Department of Astronomy and Atmospheric Sciences, Kyungpook National University, Daegu, Republic of Korea}
\email[show]{dhh13@knu.ac.kr}

\author[orcid=0000-0003-1544-8556]{Myeong-Gu Park}
\affiliation{Department of Astronomy and Atmospheric Sciences, Kyungpook National University, Daegu, Republic of Korea}
\email[show]{mgp@knu.ac.kr}

\begin{abstract}

The properties of accretion flows are affected by the angular momentum of the accreting gas. \cite{Park2009} found that the mass accretion rate, specifically, decreases significantly as the gas angular momentum increases. However, \cite{NF2011} found the decrease modest. We investigate global solutions for rotating polytropic flows in a much wider parameter space to understand their general properties within the slim disk approximation and a viscosity description suitable for both low- and high-angular momentum flows. We find that the mass accretion rate for flows with a small Bondi radius decreases steeply as the gas angular momentum increases, while for those with a large Bondi radius, it decreases gradually. Therefore, the decrease of mass accretion rate due to gas rotation can be significant or mild depending on the Bondi radius. We further investigate global solutions of accretion with outflows using the ADIOS model of \cite{BB1999}. Stronger outflows in general slightly increase the mass inflow rate at the outer boundary, but the actual mass accreted into the black hole decreases by orders of magnitude. Stronger outflows also weaken the dependence of the mass accretion rate on the gas angular momentum when the viscosity parameter \(\alpha\) is small. The intricate dependence of the mass inflow rate at the outer boundary and the mass accretion rate into the black hole on gas angular momentum will have interesting implications for the growth of black holes and their energy output.

\end{abstract}



\section{Introduction}
\label{sec:intro}

Black holes in general produce energy by accretion. Accretion converts the potential energy of matter, usually gas, to radiative or kinetic energy. Most energetic phenomena in the universe are powered by this process. The amount of radiative energy emitted is often simply parameterized by the radiation efficiency times the mass accretion rate. The radiation efficiency of a given accretion flow is the result of the physical state of the flow and the detailed microphysical processes within the accretion flow, some of which are determined by the physical conditions of accreted matter at the outer boundary. The rate of accreting mass is also determined by the outer boundary conditions of the accreted gas. 

Black holes increase their mass by accretion. The amount of mass accreted determines how fast a black hole can grow. \cite{Bondi1952} has done a pioneering study on the physical state of the accretion flow and the mass accretion rate. He solved the hydrodynamic equation for a spherical, polytropic accretion flow with no rotation. He showed that the mass accretion rate is determined by the density and temperature of the gas at infinity. Such flow has three types of solutions for a given boundary condition: subsonic, transonic, or unphysical. The transonic solution describes the transonic accretion flow, changing from subsonic infall at large radii to supersonic one at smaller radii. This transition occurs at a certain critical radius. The interesting fact is that the mass accretion rate is different for different types of solutions for the same boundary conditions. 

The transonic solution has the maximum mass accretion rate under given boundary conditions, i.e., gas density and temperature. This accretion rate is called the Bondi accretion rate. The transonic solution is possible because seemingly singular Euler equation remains regular at the critical radius. The Bondi accretion rate can be viewed as the eigenvalue of the problem determined by the regularity condition at the critical point. Since the accretion flow has to be supersonic at the black hole horizon, the accretion onto the black hole has to be transonic and proceed at the Bondi mass accretion rate.

The Bondi accretion rate is widely used in all kinds of accretion related phenomena, perhaps too widely. Given the gas conditions at infinity, such as the density \(\rho_\infty\) and the isothermal sound speed \(c_{s,\infty}\), the Bondi accretion rate onto a compact object of mass \(M\) is given by
\begin{equation}
    \MdotB =4\pi \lambda_\mathrm{B} \rB^2\rho_\infty c_{s,\infty}=4\pi\lambda_\mathrm{B}\left(GM\right)^2\rho_\infty c_{s,\infty}^{-3},
	\label{eq:mdotB}
\end{equation}
where \(M\) is the mass of the black hole, \(\rB\) the Bondi radius, defined as \(\rB=GM c_{s,\infty}^{-2} = (\rs/2) (c/c_{s,\infty})^2\), \(\rs \equiv 2GM/c^2\) the Schwarzschild radius, \(G\) the gravitational constant, \(c\) the speed of light. The flow is pressure-dominated outside the Bondi radius whereas it is gravity-dominated inside. The exact value of the dimensionless constant \(\lambda_\mathrm{B}\) depends on the adiabatic index of the gas.

In real astrophysical conditions, however, the external gas accreting onto a compact object is likely to have some angular momentum. The angular momentum of the flow fundamentally affects the nature of the accretion flow. When matter starts to be accreted with no angular momentum, the flow would be freely falling and spherical with a very low radiative efficiency \citep{Shapiro1973,Park1990a,Park1990b}. When matter has angular momentum close to the Keplerian value, the flow would assume a disk shape with minimal radial motion, resulting in a very high radiative efficiency \citep{SS1973,LP1974}. Advection-dominated accretion flow (ADAF) is somewhere between these two modes of accretion: sub-Keplerian angular momentum allows for significant radial velocity and, therefore, results in a low radiative efficiency \citep{NY1994}. Extensive studies have shown that accretion flows with different conditions of matter at the outer boundary result in different physical states of accretion and radiative outputs (see e.g., \citealt{YN2014} for a comprehensive review).

The change in radial velocity due to angular momentum inevitably affects the mass accretion rate as well: the mass accretion rate is determined by the radial velocity which is affected by the amount of the rotational motion. Strictly Keplerian rotation means no radial motion, and therefore zero mass accretion rate. Accretion flows with significant radial motion such as spherical accretion or ADAF would be expected to have varying mass accretion rate that depends on the boundary conditions of the gas.
Park (\citeyear[P09 hereafter]{Park2009}) studied the properties of hot rotating accretion flow onto compact objects for given gas conditions at the outer boundary, i.e., density, temperature, and angular momentum, within the slim disk approximation. He found out that the low angular momentum, compared to the Keplerian value, flow resembles spherical Bondi flow with a large critical radius while the high angular momentum flow resembles hot ADAF with a large critical radius. 

Moreover, the mass accretion rate depends on the value of the angular momentum of the gas being accreted, which can be significantly smaller than the classical Bondi accretion rate \(\MdotB\), for the same density and temperature of gas at the outer boundary. The mass accretion rate of the transonic solutions for the same density and temperature, but with varying amount of angular momentum of gas at the outer boundary showed that the dependence of the dimensionless mass accretion rate (\(\mdot\)), in units of corresponding (i.e., same gas density and temperature at the boundary) Bondi accretion rate, on the angular momentum of accreted gas is simply
\begin{equation}
\mdot \equiv \frac{\dot{M}}{\dot{M}_B} \simeq 9 \alpha \lambda^{-1},
\label{eq:P09}
\end{equation}
where \(\alpha\) is the usual viscosity parameter of \citet{SS1973} and 
\(\lambda \equiv \Jout/\JB\) the specific gas angular momentum at the Bondi radius \(\Jout\) in units of the Keplerian angular momentum at the Bondi radius \(\JB \equiv \sqrt{GM \rB}\). Keplerian disks have \(\lambda \simeq 1\) and non-rotating spherical Bondi flows \(\lambda = 0\). 
Although the exact value of \( \mdot\) was not calculated for small enough angular momentum flow in P09, the mass accretion rate is expected to reach the Bondi accretion rate for \(\lambda < 9.0 \alpha\).
Equation (\ref{eq:P09}) shows that the steady-state mass accretion rate is small if the flow starts with large angular momentum, and increases as the flow starts with smaller angular momentum. 

Narayan \& Fabian (\citeyear{NF2011}, NF11 hereafter) subsequently studied slowly rotating ADAF to estimate the rate at which mass accretes onto a SMBH in the nucleus of a galaxy, to understand the tight coupling between the jet power in elliptical galaxies and the mass accretion rate. They used the same slim disk approximation and \(\alpha\) viscosity prescription, but explored flows with much larger Bondi radius, i.e., gas temperature at the outer boundary with \(T_\mathrm{out} \sim 10^{6} K\) or equivalently \(\rB \sim 5 \times 10^5 \rs\), as opposed to \(\rB \sim 10^3 \rs\) in P09. Their results qualitatively agreed with those of P09 in the sense that the mass accretion rate increases as the angular momentum decreases, reaching the Bondi accretion rate for small enough gas angular momentum. However, the effect of gas angular momentum on the mass accretion rate was modest: only a factor of 3 decrease in the mass accretion rate\footnote{NF11 defined the dimensionless angular momentum as \(\mathcal{L} \equiv l_\mathrm{out}/l_\mathrm{ms}\), where the specific angular momentum of the marginally stable orbit \(l_\mathrm{ms} = \sqrt{27/8}\rs c\)}. Part of this discrepancy is due to the different value of \(\alpha\) adopted.

Another more notable discrepancy was the slope of the log-relation between the mass accretion rate and the angular momentum: the log-slope in NF11 was much shallower than that in P09. Consequently, the critical gas angular momentum at which the mass accretion rate reaches the Bondi accretion rate was different: in NF11, the critical angular momentum was \(\lambda_c \sim 0.01\), whereas in P09 \(\lambda_c \sim 9 \alpha\), both for \(\alpha = 0.1\). NF11 suspected that such discrepancy may be caused by the different parameters explored, especially, different value of the Bondi radius \(\rB\). They suggested a thorough study of how the mass accretion rate depends on viscosity parameter, angular momentum, and Bondi radius.

\cite{PH2018} extended P09 to a larger Bondi radius of \(\rB\sim 10^5\rs\), and confirmed the same dependence of the mass accretion rate on the viscosity parameter and the same \(\mdot\) -- \(\lambda\) relation discovered in P09. However, one very important difference between P09 and NF11 is the viscosity prescription: P09 adopted integrated angular momentum flux directly proportional to pressure \citep{Ab1988} while NF11 adopted that proportional to angular shear \citep{Narayan1997}. The former implementation is simple and convenient: the angular momentum equation does not require integration, and the no-torque condition at the black hole horizon is automatically satisfied. However, such implementation may overestimate viscous stress when the rotational velocity of the accretion flow is significantly lower than the Keplerian value. This implies that P09’s calculations for low-angular momentum flow may not be accurate, which may have contributed to the discrepancy with NF11 results.

One additional important aspect of hot accretion flows, such as ADAF, is outflow. \cite{NY1994} themselves realized that the ADAFs have a positive Bernoulli constant and are, therefore, prone to develop outflows. Outflows may develop in hot accretion flow due to hydrodynamic processes (e.g., \citealt{NY1994, XC1997, RCM97}; \citealt{BB1999}, BB99 hereafter; \citealt{TD2000}), magnetohydrodynamic processes \citep{Narayan2012}, or radiative preheating \citep{PO1999,PO2001}. 

Although the proper treatment of outflows requires careful multi-dimensional hydrodynamic or magnetohydrodynamic calculations, BB99 proposed simple, self-similar adiabatic inflow-outflow solutions (ADIOS) that can approximate diverse gas-dynamical or magnetically dominated winds. They assumed that the mass inflow rate is a power law of the radius, \(\mdot \propto r^s\), where the exponent \(s\) corresponds to specific cases of outflows. This radius-dependent mass inflow rate provides a good approximation to various numerical simulations \citep{Stone1999,SP2001,Begelman2012,Li2013,Yuan2012a,Yuan2012b,Yuan2015, Guo2023, Guo2024}. Observations of Sgr A\(^*\) and M87 also support the inward decrease of the mass accretion ﬂow \citep{Marrone2007, Wang2013, Kuo2014, Bower2018}. From \emph{Chandra} observations of Sgr A\(^*\), the mass inflow rate at the Bondi radius is estimated to be \(\MdotB \sim 10^{-5} \mathrm{M}_\odot \mathrm{yr}^{-1}\) \citep{Baganoff2003}. \cite{Marrone2007} determined the Faraday rotation measure in Sgr A\(^*\) at millimeter/sub-millimeter wavelengths and constrained the mass accretion rate to be in the range of \(2\times10^{-9} < \dot{M} < 2\times10^{-7}\) (in units of \(\mathrm{M}_\odot \mathrm{yr}^{-1}\)). \cite{Kuo2014} estimated constraints on the Faraday rotation measure at sub-millimeter wavelengths for M87 to be \((7.5-3.4) \times 10^5 \mathrm{rad} \; \mathrm{m}^{-2}\). They limited the mass accretion rate at 21 \(\rs\) to \( 9.2 \times 10^{-4} \mathrm{M}_\odot \mathrm{yr}^{-1}\) (corresponding to \(7.4 \times 10^{-3} \MdotB\)), which is two orders of magnitude smaller than the Bondi accretion rate onto M87. \cite{EHT2021} estimated the mass accretion rate from polarimetric observations of M87 close to the horizon as \(\Mdot \simeq (3-20) \times 10^{-4} \mathrm{M}_\odot \mathrm{yr}^{-1}\), which is consistent with that of \cite{Kuo2014}.

Recent numerical simulations also show that less than a percent of the mass inflow rate at large radii may eventually be accreted into the black hole \citep{Li2013,Inayoshi2018, BY2019, Yoon2020, Guo2023, Guo2024}. The huge difference between the mass inflow rate at the Bondi radius and the mass accretion rate at the black hole horizon will greatly affect, for given physical conditions of gas surrounding the black hole, how much energy is produced and how fast a black hole grows in mass. The combined effect of the decrease in mass accretion rate due to angular momentum and outflows will even more severely affect black hole growth and energy production. 

In this work, we investigate rotating viscous hot accretion flows onto a black hole, with or without outflow, in a larger parameter space of the viscosity parameter \(\alpha\), angular momentum at the outer boundary \(\Jout\), and Bondi radius \(\rB\). We adopt a more general viscosity prescription, as used in NF11, and assume a polytropic equation of state as in \cite{Bondi1952}. We call this flow \textquoteleft generalized Bondi flow\textquoteright, and compare the solutions with the original Bondi solutions and those from previous studies (P09; NF11). We also quantify the dependence of the mass accretion rate on \(\alpha\), \(\Jout\), and \(\rB\). 

In Section \ref{sec:EqBC}, we describe the relevant equations, boundary conditions, and method of calculations. The accretion flow solutions and their properties are addressed in Section \ref{sec:raccflow}. Solutions with outflows are explained in Section \ref{sec:ADIOS}. Finally, we summarize our results in Section \ref{sec:summary}.

Throughout this work, we use \(\mdot \equiv \dot{M} / \MdotB\) as the dimensionless mass accretion rate (in units of the Bondi accretion rate) and \(\lambda\equiv \Jout/\JB\) as the dimensionless gas angular momentum at the outer boundary \(\Jout=\Omega(\rout)r^2_\mathrm{out}\) in units of the Keplerian angular momentum at the Bondi radius \(\JB\). The accreting gas is assumed to be composed of pure hydrogen and fully ionized. Ions and electrons are assumed to have the same temperature.

\section{Equations and Method of Integration}
\label{sec:EqBC}

\subsection{Equations}
\label{sec:eq} 

We adopt slim disk approximation \citep{Ab1988}, which allows for the radial motion of the flow and the critical point, essential for determining the proper global solution and the mass accretion rate (P09). We assume steady-state accretion. In the slim disk approximation, all physical quantities are functions of radius (\(r\)) only, and the accretion flow is described by four equations: the continuity equation, the radial momentum equation, the angular momentum equation, and the energy equation.

The continuity equation relates the mass accretion rate \(\dot{M}\) to the gas density \(\rho\) and the radial velocity \(v_r\) at radius \(r\),
\begin{equation}
    \dot{M}=-4\pi r^2\rho v_r .
	\label{eq:continuity}
\end{equation}
The velocity \(v_r\) is taken to be negative for inflow. We set the disk scale height equal to the radius, \(H = r\), so that the solutions can be readily compared with spherical Bondi solutions (NF11). This approximation is valid only for hot accretion flows.

We adopt the Paczy\'{n}ski-Wiita potential to mimic relativistic effects \citep{PW1980}, \(\Phi_\mathrm{PW}=-(GM)/(r-\rs)\). The Keplerian angular velocity in the Paczy\'{n}ski \& Wiita potential is \(\Omega^2_K(r) = GMr^{-1}(r-\rs)^{-2}\).

The radial momentum equation in the presence of rotation and pressure is
\begin{equation}
    v_r\dfrac{dv_r}{dr} + (\Omega_K^2 - \Omega^2)r + \dfrac{1}{\rho}\dfrac{dP}{dr} = 0,
	\label{eq:rmom}
\end{equation}
where \(\Omega(r)\) is the angular velocity at radius \(r\). The gas pressure \(P(r)\) and gas density \(\rho(r)\) are related by \(P = \rho c_s^2\), where \(c_s\) is the isothermal sound speed. 

The integrated form of the angular momentum equation in the slim disk approximation is given by
\begin{equation}
    \rho v_r(\Omega r^2 - \JO) = \eta\alpha rP,
	\label{eq:angmom}
\end{equation}
where \(\alpha\) is the viscosity parameter from \cite{SS1973}. The integration constant \(\JO\) is the specific angular momentum accreted into the black hole, and has to be determined along with the solution. The prescription for the parameter \(\eta\) depends on the detailed implementation of the viscosity description (P09): \(\eta\) can range from \(\eta = -2\) \citep{Ab1988}, \(\eta = -1\) \citep{Nakamura1997}, \(\eta = (r/\Omega_K)(d\Omega/dr)\) \citep{Narayan1997}, to \(\eta = (r/\Omega_K)(d\Omega_K/dr)\) \citep{Yuan1999}. P09 chose \(\eta = -1\) because such choice simplifies the angular momentum equation into an algebraic one and automatically satisfies the no-torque condition at the black hole horizon \citep{Ab1988,Yuan2000}. However, this prescription may overestimate the shear stress when the flow is sub-Keplerian, such as in low-angular momentum flow, or when the shear of the rotation is minimal, such as at the radius outside the Bondi radius. Therefore, we adopt a viscosity prescription that is proportional to the angular shear (NF11),
\begin{equation}
 \eta= \frac{r^2}{c_s} \frac{d\Omega}{dr},
 \label{eq:eta}
\end{equation}
and the angular momentum equation becomes
\begin{equation}
 \frac{d\Omega}{dr} = \frac{v_r}{\alpha r^3 c_s} 
                    \left( \Omega r^2 - \JO \right) .
 \label{eq:dOmegadr}
\end{equation}
This prescription provides a better description of the viscous flow over larger parameter space and also facilitates comparison with the results of NF11.

For consistent comparison with the Bondi solutions, which assume polytropic equation of state \citep{Bondi1952}, we also assume polytropic equation of state
\begin{equation}\label{eq:poly}
    \frac{P}{P_o} = \left( \frac{\rho}{\rho_o} \right)^\gamma,
\end{equation}
where \(P_o\) and \(\rho_o\) are the pressure and density at a given radius, such as the outer boundary. This equation replaces the energy equation. When the accretion flow is hot and advective, radiative cooling is not important compared to energy advection through radial infall as in spherical Bondi flow (\citealt{NY1994}; P09; NF11), and the polytropic equation of state is a good approximation. We also assume fully ionized, one-temperature hydrogen gas, with \(c_s^2 = P / \rho = \mu^{-1} k T m_p^{-1}\) where \(T\) is the gas temperature, \(k\) is the Boltzmann constant, \(m_p\) is the proton mass, and \(\mu = 1/2\) is the molecular weight.

\subsection{Method of Integration}

As noted by \cite{Bondi1952}, transonic accretion flow without rotation is an eigenvalue problem in the sense that, for given boundary conditions (gas density and temperature), only the flow with a specific mass accretion rate (or equivalently, with a specific radial velocity at the outer boundary) can pass through the critical (a.k.a. sonic) point and become supersonic. Accretion flows with rotation in the slim disk approximation exhibit the same behavior: for given density, temperature, and specific angular momentum of the gas at the outer boundary, the mass accretion rate must be a specific critical value to become transonic (\citealt{Ab1988}; P09). 

The radial momentum equation (\ref{eq:rmom}) can be rearranged into a critical form by utilizing the continuity equation (\ref{eq:continuity}) and the polytropic equation of state (\ref{eq:poly}),
\begin{equation}\label{eq:critform}
  \frac{d v_r}{dr} 
  = - \frac{ ( \Omega_K^2 - \Omega^2 ) r v_r 
             - 2 \gamma c_s^2 r^{-1} v_r  }
           { v_r^2 - \gamma c_s^2 } .
\end{equation}
Direct integration of equation (\ref{eq:critform}), from the outer boundary across the critical point \(r_c\), where \( v_r = \gamma^{1/2} c_s \), is not possible because the denominator approaches zero close to the critical point, and the equation diverges. A regular solution is possible only if the numerator also becomes zero at the critical point. This regularity condition is met only when the mass accretion rate is precisely equal to the critical value. 

Regular solutions with a critical point for given outer boundary conditions can typically be constructed using the shooting method (\citealt{Park1990a,Park1990b}; P09; \citealt{Yuan1999,Yuan2000}) or the relaxation method (\citealt{Narayan1997}; NF11). In the shooting method, the equations are integrated inward, starting from the outer boundary with guessed initial values for the mass accretion rate. Such an initial integration yields a subsonic or diverging solution as the integration proceeds close to the critical point, and fails. Iterated adjustments of the mass accretion rate enable the integration to proceed ever closer to the critical point with regular behavior. When the integration reaches close enough to the critical point, where \( v_r \approx \gamma^{1/2} c_s \), the solutions are extrapolated across the critical point to integrate inward down to the inner boundary. This approach is simple and bears some resemblance to the actual physical behavior of the accretion flow. Although it is quite suitable for accretion flows with a small enough outer boundary radius (as in P09), the required adjustments become excessively sensitive for a large outer boundary radius (as in NF11). Since we aim to explore a wide range of parameter space, including the outer boundary radius, the relaxation method is better suited (NF11). 

We need to solve two differential equations (\ref{eq:dOmegadr}) and (\ref{eq:critform}) while simultaneously determining the mass accretion rate \(\dot{M}\), the angular momentum at the innermost stable circular orbit (ISCO) \(\JO\), and the critical radius \(\rcr\), all for given outer boundary conditions at \(r = \rout\), i.e., 
\begin{equation}
    \rho = \rho_\mathrm{out}, \;\cs = c_\mathrm{s,out}, \;\Omega = \Omega_\mathrm{out}.
\end{equation}
The regularity condition at the critical point provides two additional boundary conditions:
\begin{equation}
    -\left(\Omega_\mathrm{K}^2 - \Omega^2 \right)r v_r + 2\gamma \cs^2 \dfrac{v_r}{r}\Bigg|_{r=\rcr} = 0,
\end{equation}
and
\begin{equation}
    v_r^2 - \gamma \cs^2 \Big|_{r=\rcr} = 0.
\end{equation}

In our calculations, we adopt the no-torque condition at the horizon and the assumption that angular momentum transport is minimal in the supersonic part of the flow, as adopted in NF11. As a consequence, the specific angular momentum remains constant for \(r < \rcr\) and the flow has to satisfy \(d(\Omega r^2)/dr = 0\) as an additional boundary condition at \(r=\rcr\) (NF11):
\begin{equation}
    \Omega \rcr^2 -\JO = -\dfrac{2\alpha c_s \Omega \rcr^2}{v_r}.
\end{equation} 

The iterative integration proceeds as follows. We first choose the adiabatic index \(\gamma\), the viscosity parameter \(\alpha\), the outer radius \(\rout\), and the angular momentum at the outer boundary \(\lambda\). We assume a trial value for \(\rcr\) and integrate equations (\ref{eq:dOmegadr}) and (\ref{eq:critform}) from \(\rout\) to \(\rcr\) using the relaxation method. When the integrated solution diverges due to an incorrect trial value for \(\rcr\), we adjust \(\rcr\) and repeat the relaxation procedure until the solution converges. The mass accretion rate \(\mdot\) and the specific angular momentum \(\JO\) are determined self-consistently through relaxation. Once \(\rcr\) is determined, we apply L'H\(\hat{\mathrm{o}}\)pital's rule to pass the critical point and then integrate inward down to the inner boundary with a usual integration driver. 

In this work, we choose the adiabatic index of the accreting gas to be \(\gamma=5/3\). The density is expressed in units of \(\rho_\mathrm{0}=m_p/(\sigma_T \rs)\). We set the outer boundary at the Bondi radius, \(\rout=\rB\), the temperature at \(\rout\) is \(T_\mathrm{out} = \mu m_p c_{s,\infty}^2 / k = (\mu/2)(m_p c^2 / k) (\rB / \rs)^{-1} = 2.73 \times 10^{12} (\rB / \rs )^{-1} \, \mathrm{K}\), and the inner boundary at the innermost stable circular orbit, \(r_\mathrm{in} = r_\mathrm{ISCO} = 3 \rs\).

\begin{figure} 
    \centering
    \includegraphics[width=0.8\columnwidth]{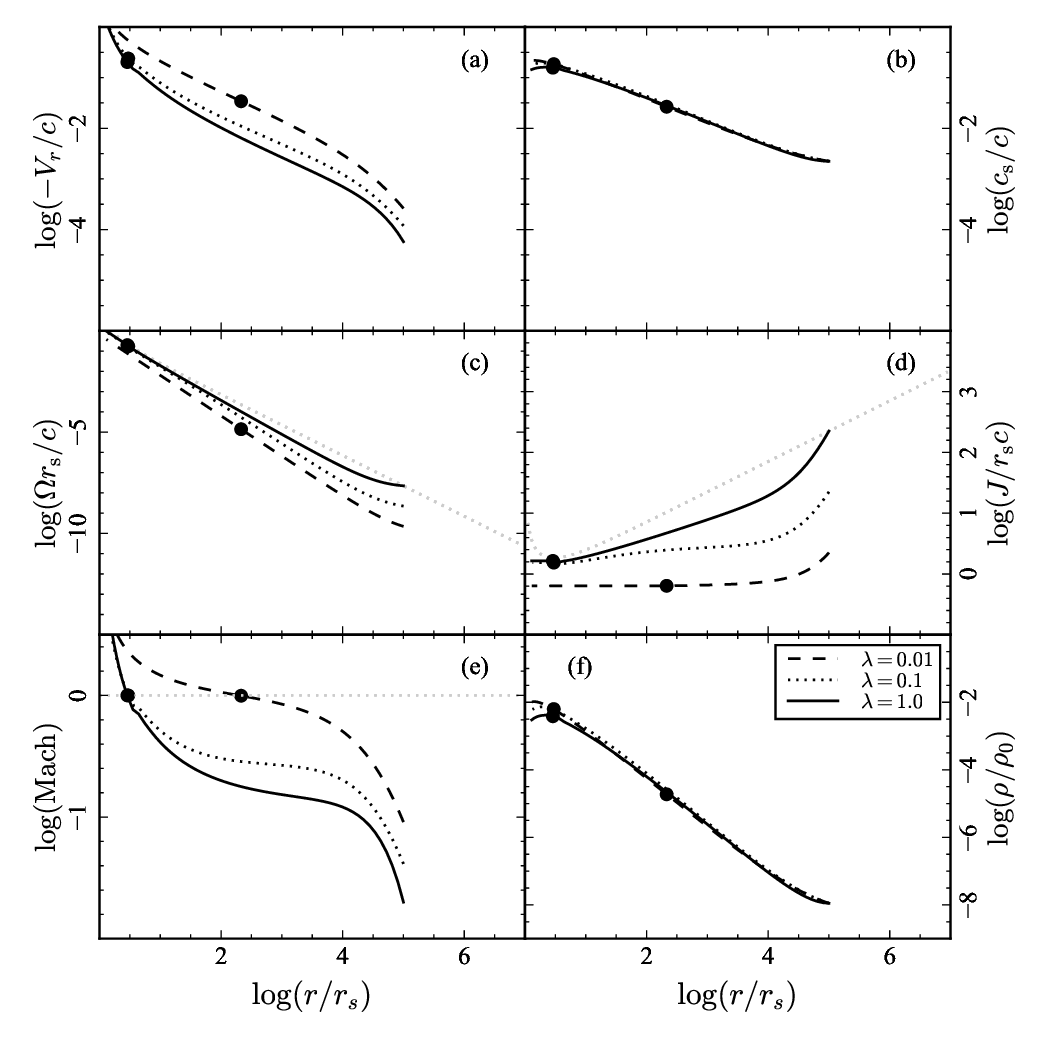}
    \caption{Flow profiles of accretion flows with \(\gamma = 5/3\), \(\alpha = 0.1\), at a Bondi radius of \(\rB = 1.0 \times 10^5 \rs\), corresponding to \(T_\mathrm{out} = 2.7 \times 10^{7} \mathrm{K}\) or \(\cs = 2.2 \times 10^{-3} c\). Different line styles denote different values of the angular momentum: \(\lambda=1.0\) for high angular momentum (solid line), \(\lambda=0.1\) for intermediate angular momentum (dotted line), and \(\lambda = 0.01\) for low angular momentum (dashed line). The grey dashed lines in the panels denote the Keplerian angular velocity (c) and angular momentum (d). The filled circles denote the positions of the critical points.}
    \label{fig:profiles}
\end{figure}

\section{Generalized Bondi Flow}
\label{sec:raccflow}

\subsection{Flow Properties}
\label{sec:profile}

First, we discuss the general properties of the accretion flow solutions we have found. Figure \ref{fig:profiles} shows typical solutions for accretion flows with a viscosity parameter \(\alpha = 0.1\). The outer boundary \(\rB = 10^5 \rs\) corresponds to \(\cs = 2.2 \times 10^{-3} c\) or \(T_\mathrm{out} = 2.5 \times 10^{-6} \, m_p c^2 k^{-1} = 2.7 \times 10^{7} \mathrm{K}\). The three different lines in each panel correspond to varying angular momentum at the outer boundary \(\rB\): solid lines for a high angular momentum flow (\(\lambda = 1.0\)), dotted lines for an intermediate angular momentum flow (\(\lambda = 0.1\)), and dashed lines for a low angular momentum flow (\(\lambda = 0.01\)).

The \(\lambda = 1.0\) high angular momentum flow has Keplerian angular momentum at the outer boundary (\(\Jout = \JB\)) and, therefore, accretes slowly into the black hole (solid lines in Figure \ref{fig:profiles}a). Beyond the Bondi radius, the gas rotates as a rigid body because viscosity wins weak gravity at large radii. Inside \(\rB\), viscous shear driven by gravity transports angular momentum outward, but keeps rotation (solid line in Figure \ref{fig:profiles}c) close to the Keplerian value (dotted diagonal line in Figure \ref{fig:profiles}c). The transport of angular momentum is rather efficient down to \(\sim 10^4 \rs\), inside which it becomes less efficient (Figure \ref{fig:profiles}d).

The gradient of the radial infall velocity increases gradually as the flow approaches the critical point (filled circles in Figure \ref{fig:profiles}). The flow passes the critical point at \(\rcr = 3.04 \rs\), which is quite close to the innermost stable circular orbit, \(r_\mathrm{ISCO} = 3r_\mathrm{s}\). Since the flow rotation is nearly Keplerian, the radial velocity is much smaller than the free-fall value, and the mass accretion rate is only 20\(\%\) of the Bondi rate (\(\mdot = 0.2\)) for this specific flow solution. 

The angular momentum profile of the \(\lambda = 0.01\) low angular momentum accretion flow (dashed lines in Figure \ref{fig:profiles}) is significantly different from that of the high angular momentum flow (Figure \ref{fig:profiles}d). The angular momentum is efficiently removed between \(\rB\) and \(\sim 10^4 \rs\) but remains nearly constant inside \(\sim 10^4 \rs\) (dashed line), unlike that in the high angular momentum flow (solid line). 

The radial velocity of the \(\lambda = 0.01\) flow is correspondingly larger than that of the high angular momentum accretion flow. The profiles of the low angular momentum flow (dashed lines in Figure \ref{fig:profiles}) resemble those of non-rotating Bondi flow. It has a critical point at a radius \(\rcr = 187.6 \rs\), much larger than \(\rcr = 3.04 \rs\) in the high angular momentum flow with \(\lambda = 1.0\). Most importantly, the mass accretion rate is almost equal to the Bondi accretion rate (\(\mdot \simeq 1\)). 

The \(\lambda=0.1\) intermediate angular momentum flow shows profiles in between those of the high and low angular momentum flows (dotted lines in Figure \ref{fig:profiles}). The intermediate angular momentum flow passes through the critical point at \(\rcr = 3.14 \rs\) with approximately half the Bondi accretion rate (\(\mdot = 0.47\)).

The rotating Bondi flow in this work shows nearly the same properties as the accretion flows studied in P09 and NF11, despite the difference in the equation of state: a polytropic equation of state in this work versus ideal gas with cooling and heating in P09 and NF11. 
The critical point forms at successively smaller radii, and the radial infall velocity becomes a smaller fraction of the free-fall velocity as the angular momentum of the gas increases. Flows with larger angular momentum become typical disk flow while those with smaller angular momentum resemble spherical accretion flows. The temperature of the flow is a factor of a few lower than the virial temperature, with \( T(r) \propto r^{-1}\), regardless of the gas angular momentum.

\subsection{Mass Accretion Rate}
\label{sec:mdot}

The mass accretion rate of rotating viscous accretion flows depends on angular momentum, in addition to the density and temperature of the accreting gas at the outer boundary (P09; NF11; \citealt{PH2018}). P09 showed that the mass accretion rate of hot, radiatively inefficient accretion flows in units of the Bondi accretion rate is inversely proportional to angular momentum and proportional to the viscosity parameter, \(\mdot \sim 9\alpha \lambda^{-1}\). Hence, the mass accretion rate of the flows with near-Keplerian angular momentum can be significantly smaller than the classic Bondi accretion rate. However, NF11, who considered slowly rotating accretion flows in the context of cooling flows in cluster of galaxies found that the log-log slope for \(\mdot\) and \(\lambda\) is approximately -0.3 for \(\alpha = 0.1\) and \(\gamma = 5/3\), as opposed to -1 in P09, and that the decrease in the mass accretion rate is rather modest. 

We searched for solutions of the accretion flow as described in \S \ref{sec:eq} in a larger parameter space than in P09 or NF11, focusing on the dependence of the mass accretion rate on the angular momentum, the viscosity parameter, and the Bondi radius. The ranges of gas temperatures at the Bondi radius and viscosity parameters explored are \(T_\mathrm{out} = 2.5 \times 10^{-6} \, m_p c^2 k^{-1} = 2.73 \times 10^6 \mathrm{K}\) to \(1.0 \times 10^{-4} \, m_p c^2 k^{-1} = 1.09 \times 10^9 \mathrm{K}\) (corresponding to Bondi radii of \(\rB = 2.5 \times 10^3 \rs\) to \(1.0 \times 10^5 \rs\)) and \(\alpha = 0.001 \sim 0.1\).

\subsubsection{Dependence on Angular Momentum}
\label{sec:depend_AM}

\begin{figure} 
    \centering
    \includegraphics[width=0.9\columnwidth]{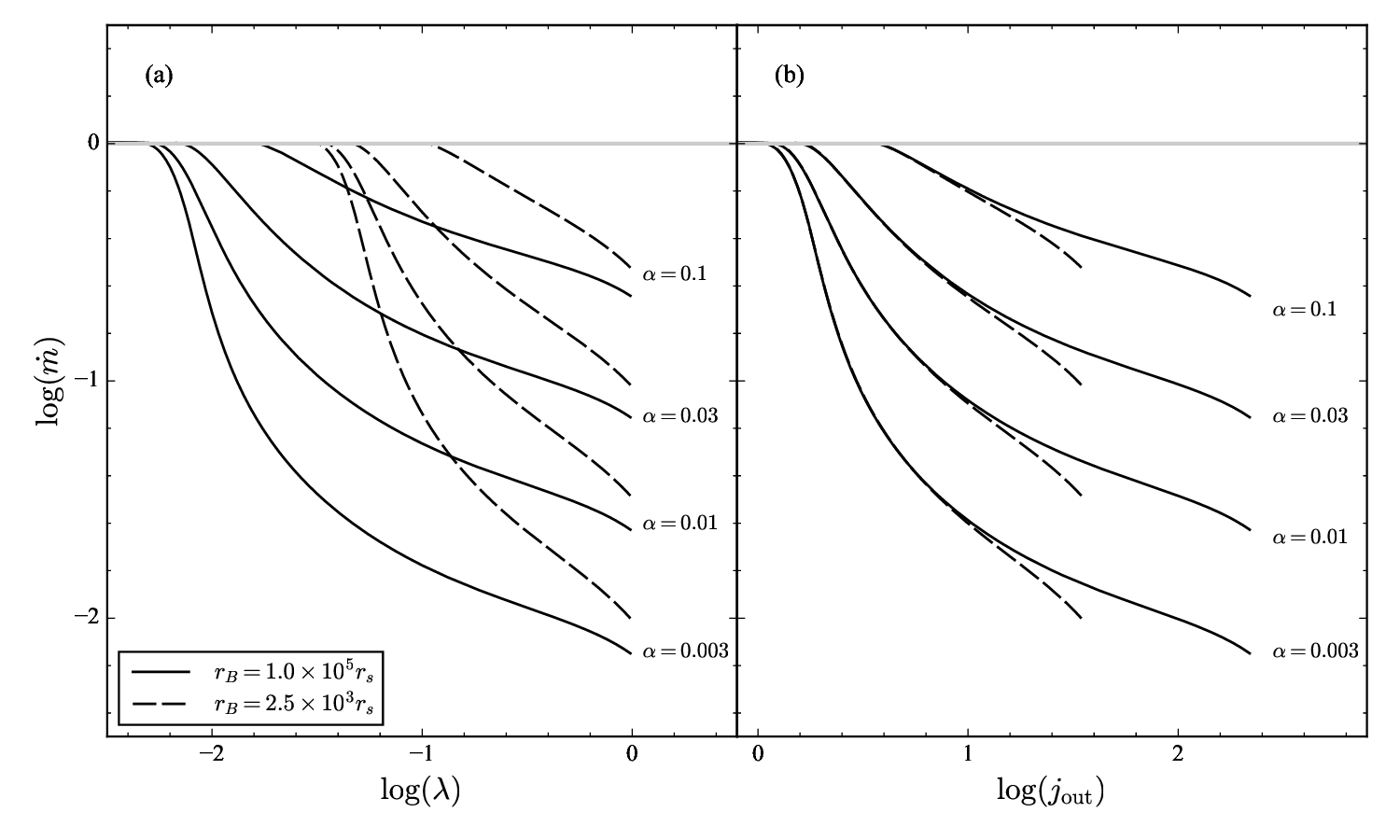}
    \caption{(a) The mass accretion rate \(\mdot\) as a function of angular momentum \(\lambda \equiv \Jout/\JB\) for four different viscosity parameters (\(\alpha=0.1, 0.03, 0.01, 0.003\)). Different line styles denote flows with different Bondi radii: solid lines for \(\rB=1.0\times 10^5\rs\) and dashed lines for \(\rB = 2.5 \times 10^3 \rs\). (b) The mass accretion rate as a function of angular momentum \(\jout \equiv \Jout / (\rs c)\).}
    \label{fig:dep_AM}
\end{figure}

Figure \ref{fig:dep_AM}a shows the mass accretion rate \(\mdot\) of the solutions for small (\(\rB = 2.5 \times 10^3 \rs\)) and large (\(\rB = 1.0 \times 10^5 \rs\)) Bondi radii as a function of the gas angular momentum at the Bondi radius (\(\lambda\)). The basic dependence of \(\mdot\) on \(\lambda\) is consistent with both P09 and NF11: \(\mdot\) decreases as \(\lambda\) increases. When the flow has near-Keplerian rotation, i.e., for large enough \(\lambda\), Equation (\ref{eq:angmom}) with \(\eta = -1\) is a good approximation, and the slope of \(\log \mdot\) versus \(\log \lambda\) is approximately -1 (dashed lines in Figure \ref{fig:dep_AM}a) for \(0.1 \lesssim \lambda \lesssim 1\), as seen in P09. However, the slope of \(\log \mdot\) versus \(\log \lambda\) for larger \(\rB\) is much shallower (solid lines in Figure \ref{fig:dep_AM}a), as seen in NF11. 

The detailed behavior of \(\mdot(\lambda)\), in general, depends on both \(\alpha\) and \(\rB\). The slope of \(\log \mdot\) versus \(\log \lambda\) is no longer constant, but changes as \(\lambda\) decreases, which is subtly seen in Figure 2 of NF11. As \(\lambda\) decreases, the roughly constant slope becomes progressively steeper, then decreases to zero as the mass accretion rate reaches the Bondi rate (\(\mdot = 1\)). The difference from P09 in small \(\lambda\) flows is not surprising because the viscosity description adopted in P09 is reasonable only for high \(\lambda\) flows. 

In all cases, the mass accretion rate reaches the Bondi accretion rate for small enough \(\lambda\). The critical value \(\lambda_{cr}\) at which this (\(\mdot = 1\)) occurs depends on the specific values of \(\alpha\) and \(\rB\) (Figure \ref{fig:dep_AM}a). The value of \(\lambda_{cr}\) is greater for larger \(\alpha\): a larger value of \(\alpha\) implies more efficient removal of angular momentum within the flow, thereby making the flow closer to non-rotating Bondi flow. Flows with a smaller value of \(\alpha\) reach the Bondi rate at smaller \(\lambda\) because angular momentum removal is less efficient. Flows with a larger value of \(\rB\) also have a smaller \(\lambda_{cr}\). Flows with larger \(\rB\) have greater angular momentum at the outer boundary for the same value of \(\lambda\) because \( \Jout = \lambda J_B \propto \rB^{1/2}\). A larger amount of angular momentum has to be removed until the flow practically becomes a Bondi flow. 

This is better shown when the angular momentum is expressed in units of \(\rs c\). Figure \ref{fig:dep_AM}b shows the same solutions as in Figure \ref{fig:dep_AM}a in \(\log \mdot\) versus \(\log \jout\) where \(\jout \equiv \Jout /(\rsc)\). The critical value \(\jcr \equiv \lcr\JB / (\rsc)\) interestingly does not depend on \(\rB\) anymore. The closeness of a certain rotating accretion flow to non-rotating Bondi flow appears to be determined by angular momentum scaled to \(\rs c\) (NF11) rather than that scaled to \(\JB\) (P09). As \(\jout\) becomes larger than \(\sim 10\), however, \(\mdot\) starts to depend on \(\rB\) (solid versus dashed lines in Figure \ref{fig:dep_AM}b). The value of \(\jcr\) still depends on the value of \(\alpha\), which specifies the efficiency of angular momentum removal: \(\jcr \simeq 3.6, 1.6, 1.2\), and \(1.1\) for \(\alpha = 0.1, 0.03, 0.01\), and \(0.003\), respectively. So it seems that the property of low angular momentum flow is better described by \(\jout\) while that of high angular momentum flow is better described by \(\lambda\).

\begin{figure} 
    \centering
    \includegraphics[width=0.9\columnwidth]{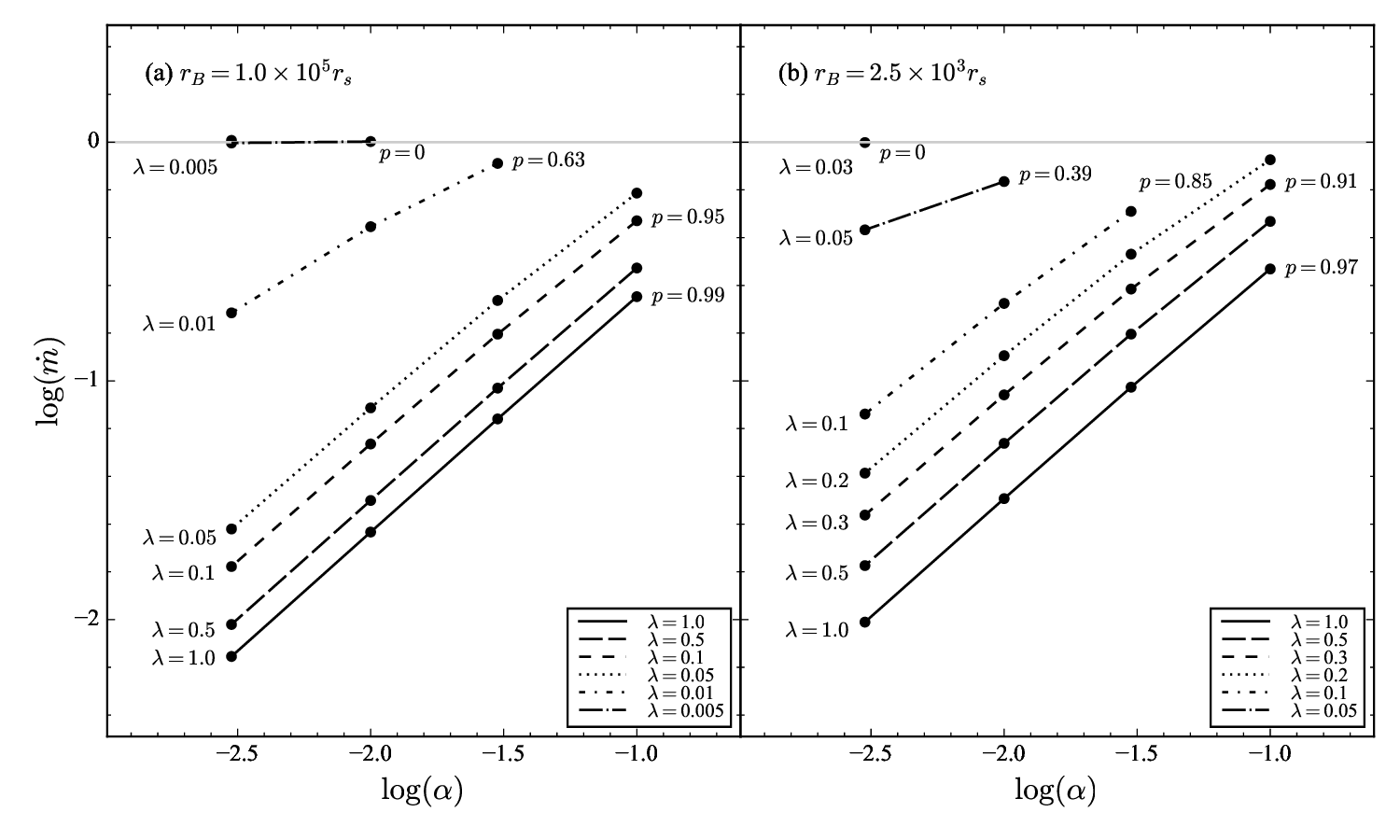}
    \caption{The mass accretion rate (\(\log (\mdot)\)) versus the viscosity parameter (\( \log(\alpha)\)) for accretion flows with a given \(\lambda\). The left and right panels are for accretion flows with \(\rB = 1.0 \times 10^5 \rs\) and \(\rB = 2.5 \times 10^3 \rs\), respectively.}
    \label{fig:mdot_alpha}
\end{figure}

\subsubsection{Dependence on the Viscosity Parameter}
\label{sec:depend_alpha}

The dependence of the mass accretion rate \(\mdot\) on the viscosity parameter \(\alpha\) is almost linear for large enough \(\lambda\). Figure \ref{fig:dep_AM} shows \(\mdot\) for different values of \(\alpha\) (0.1, 0.03, 0.01, 0.003 from top to bottom) and two values of \(\rB\) (dashed lines for \(2.5 \times 10^3 \rs\) and solid lines for \(1.0 \times 10^5 \rs\)). For \(\lambda \gtrsim 0.1\), \(\mdot \propto \alpha\) (\(d \log \mdot / d \log \alpha = 1\)) as shown in P09. Self-similar ADAFs also have mass accretion rates proportional to \(\alpha\) (Narayan \& Yi 1994). However, the dependence on \(\alpha\) becomes weaker for smaller values of \(\lambda\) (\(\lambda \lesssim 0.1\)). 

Figure \ref{fig:mdot_alpha} shows \(\log \mdot\) versus \(\log \alpha\) for various values of \(\lambda\) and two values of \(\rB\): Figure \ref{fig:mdot_alpha}a for \(\rB =1.0 \times 10^5 \rs\) and Figure \ref{fig:mdot_alpha}b for \(\rB = 2.5 \times 10^3 \rs\). It shows that the mass accretion rate \(\mdot\) of flows with large enough \(\lambda\) (\(\lambda \geq 0.05\) for \(\rB = 1.0 \times 10^5 \rs\), \(\lambda \geq 0.1\) for \(\rB = 2.5 \times 10^3 \rs\)) depends linearly on \(\alpha\), as seen in P09. If we define a power index \(p\) as \(\mdot \propto \alpha^p\), these flows have \(p \simeq 1\). However, as \(\lambda\) decreases, \(p\) also decreases. The dependence of \(\mdot\) on \(\alpha\) weakens:
\(p=0.63\) when \(\lambda=0.01\) for \(\rB=1.0 \times 10^5\rs\) and \(p=0.39\) when \(\lambda=0.03\) for \(\rB=2.5 \times 10^3\rs\).
When \(\lambda\) becomes small enough, flow becomes Bondi-like and its mass accretion rate reaches the Bondi rate (\(\mdot =1\)) with no dependence on \(\alpha\) (\(p \simeq 0\)). We list specific values of \(p\) for some values of \(\alpha\) and \(\rB\) in Figure \ref{fig:mdot_alpha}.
To summarize, the mass accretion rate of high angular momentum flows is proportional to \(\alpha\), but its dependence on \(\alpha\) decreases as angular momentum decreases until the flow becomes Bondi-like and the mass accretion rate becomes independent of \(\alpha\).

\begin{figure} 
    \centering
    \includegraphics[width=0.8\columnwidth]{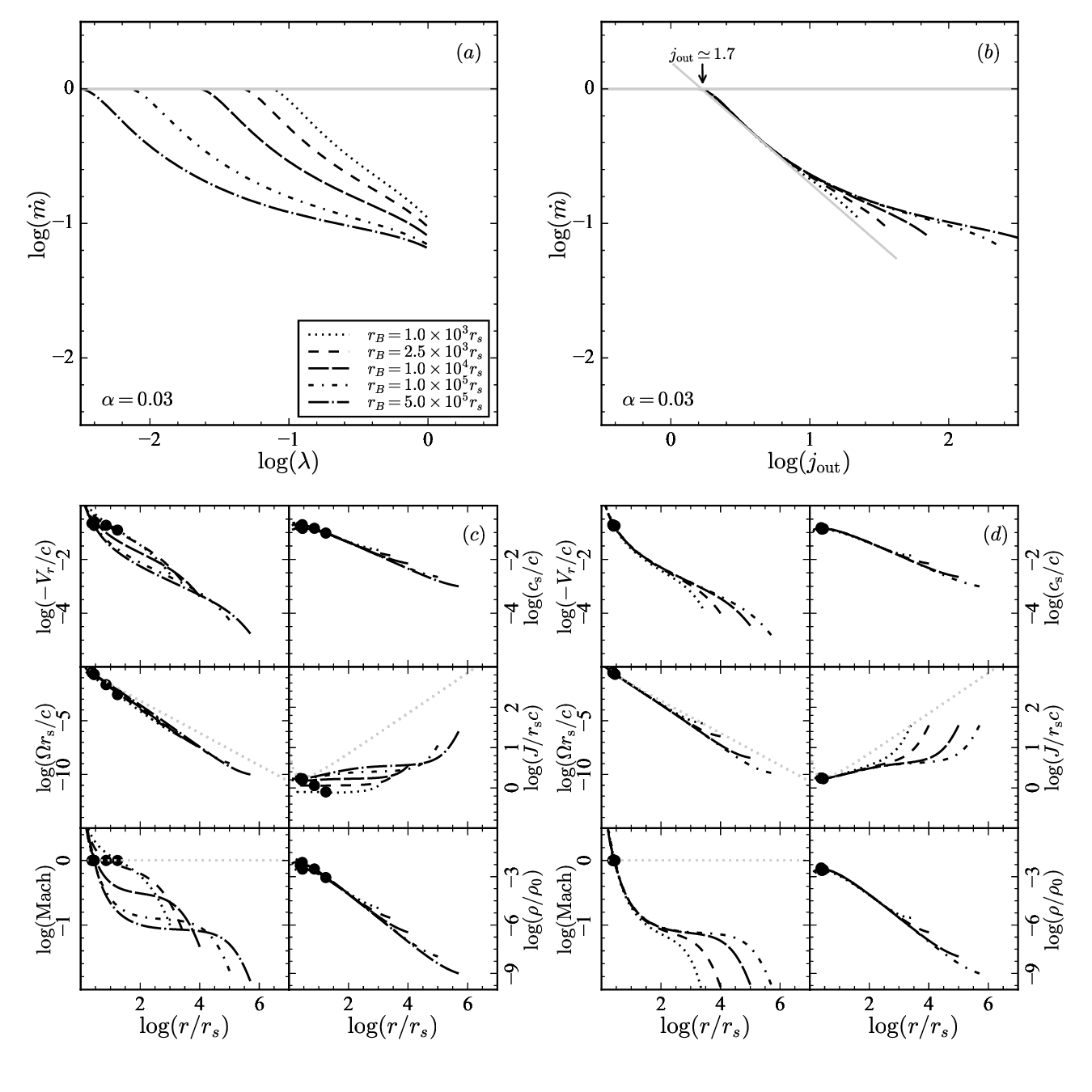}
    \caption{The mass accretion rate as a function of angular momentum parameter in (a) for \(\lambda\) and (b) for \(\jout\), for accretion flows with \(\gamma=5/3\) and \(\alpha = 0.03\) at various Bondi radii. Flow profiles for the accretion flow with specific angular momentum: (c) for \(\lambda = 0.1\) and (d) for \(\jout = 35.5\). Different line styles denote different Bondi radii: dotted lines for \(\rB = 1.0 \times 10^3 \rs\), dashed lines for \(\rB = 2.5 \times 10^3 \rs\), long-dashed lines for \(\rB = 1.0 \times 10^4 \rs\), dot-dashed lines for \(\rB = 1.0 \times 10^5 \rs\), and dot-long-dashed lines for \(\rB = 5.0 \times 10^5 \rs\). The filled circles in the bottom panels denote the positions of the critical points.}
    \label{fig:mdotrb}
\end{figure}

\subsubsection{Dependence on the Bondi Radius}
\label{sec:depend_rb}

One of the main questions unanswered in P09 and NF11 is how the Bondi radius affects flow properties.
Solutions in Figure \ref{fig:dep_AM} show that the mass accretion rate of accretion flows with a small Bondi radius (i.e., high temperature at the outer boundary) decreases significantly as angular momentum increases (long-dashed lines in Fig. \ref{fig:dep_AM}): it is described approximately by \(\mdot \sim 9 \alpha / \lambda\) for \(\lambda \geq 0.1\), agreeing well with P09. However, for accretion flows with a large Bondi radius (solid lines), the dependence of \(\mdot\) on \(\lambda\) is much shallower, which agrees with the results of NF11.

\begin{figure} 
    \centering
    \includegraphics[width=0.8\columnwidth]{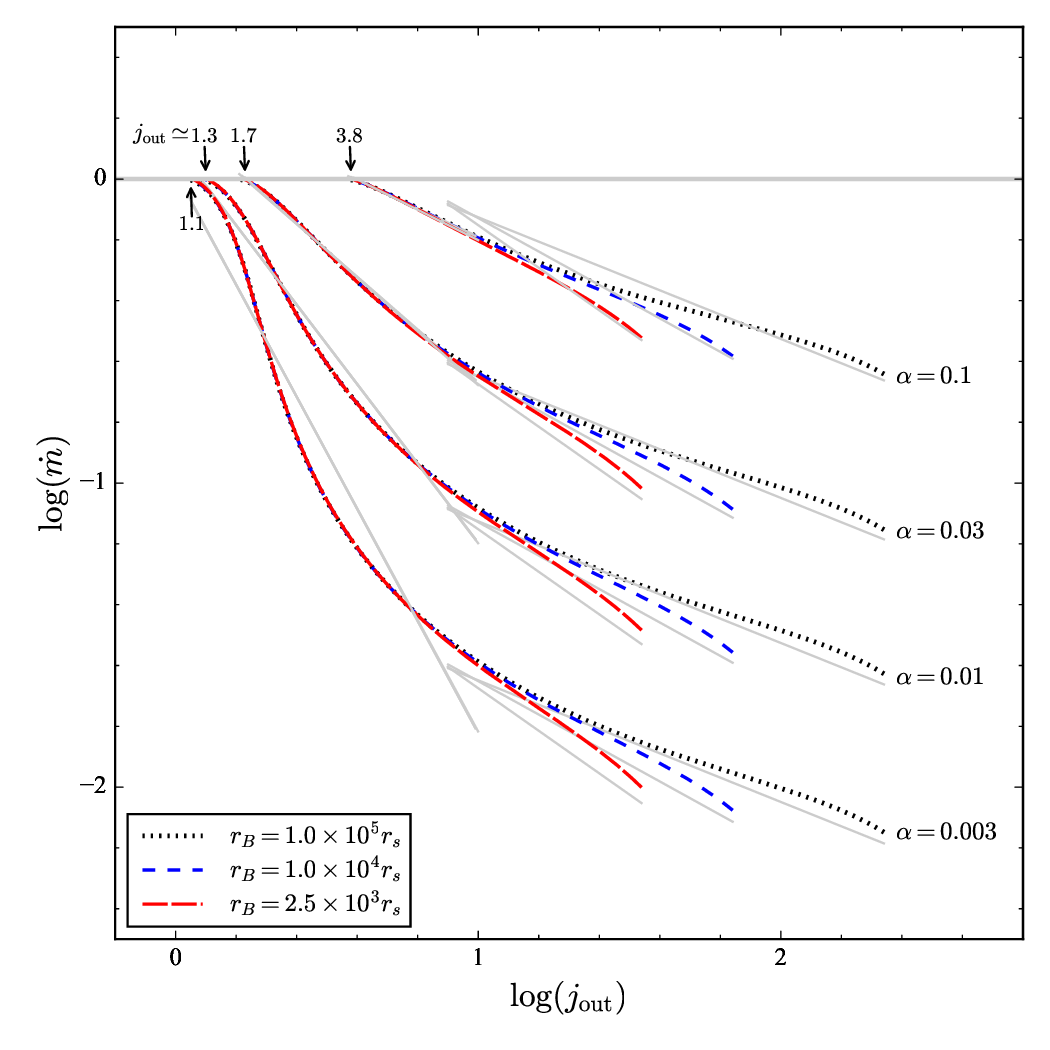}
    \caption{The mass accretion rate \(\mdot\) as a function of angular momentum \(\jout\) for accretion flows with \(\gamma=5/3\) and four values of \(\alpha\) at three Bondi radii. Different line styles denote the accretion flows at \(\rB = 1.0 \times 10^5 \rs\) (dotted lines), \(\rB = 1.0 \times 10^4 \rs\) (dashed lines), and \(\rB = 2.5 \times 10^3 \rs\) (long-dashed lines). Grey lines represent the mass accretion rates approximated by fitting Equations \ref{eq:gen_P09} and \ref{eq:eq_mdot_jout} for high- and low- angular momentum flows, respectively.}
    \label{fig:mdotrb4}
\end{figure}

Our solutions clearly show how the dependence of the mass accretion rate on \(\lambda\) changes as the Bondi radius changes.
Figures \ref{fig:mdotrb}a shows the mass accretion rate \(\mdot\) as a function of angular momentum \(\lambda\) for different values of the Bondi radius: dotted line for \(\rB = 1.0\times 10^3 \rs\), dashed line for \(\rB = 2.5 \times 10^3 \rs\), long-dashed line for \(\rB = 1.0 \times 10^4 \rs\), dot-dashed line for \(\rB = 1.0 \times 10^5 \rs\), and dot-long-dashed line for \(\rB = 5.0 \times 10^5 \rs\). The mass accretion rate \(\mdot\) depends roughly linearly on \(\lambda\) for small \(\rB\), such as \(\rB = 1.1 \times 10^3 \rs\) (dotted line) as shown in P09.
However, flows with larger \(\rB\) show that \(\mdot(\lambda)\) has an increasingly shallower slope in \(\lambda\) as \(\rB\) increases. 
Figure \ref{fig:mdotrb}c shows the flow profiles of accretion flows with the same \(\lambda\) for various Bondi radii. For a given \(\lambda\), an accretion flow with small \(\rB\) has low physical angular momentum at the outer boundary and approaches the critical point shortly after losing a small amount of angular momentum. On the other hand, an accretion flow with a larger Bondi radius has (for the same value of \(\lambda\)) higher physical angular momentum at the outer boundary. The accreting gas has to lose a larger amount of angular momentum and therefore reaches the critical point only at a much smaller radius.

Our results show that the properties of high-angular momentum flows, especially the mass accretion rate, mostly depend on the angular momentum parameter \(\lambda\) (angular momentum normalized by the Bondi angular momentum \(J_B\)), regardless of the Bondi radius, as noted by P09. However, low angular momentum flows show rather complex properties in terms of \(\lambda\). NF11 studied low angular momentum flows in terms of \(\jout \equiv J_{out} / (\rs c)\). Hence, we investigate how the dependence of flow properties on angular momentum differs when angular momentum is expressed in \(\jout\) instead of \(\lambda\).

Figure \ref{fig:mdotrb}d shows the profiles of flows with a fixed value of \(\jout =35.3\) but with different Bondi radii \(\rB\). As the accreting gas falls inside the Bondi radius, it rapidly loses angular momentum and then approaches the critical point with similar flow behavior. Profiles of the velocity, angular momentum, and Mach number all converge to a single profile at small radii, regardless of the value of \(\rB\), unlike the flows with the same value of \(\lambda\) shown in Figure \ref{fig:mdotrb}c.

Figure \ref{fig:mdotrb}b shows \(\mdot(\jout)\) for comparison with \(\mdot(\lambda)\) shown in Figure \ref{fig:mdotrb}a for the same value of \(\rB\). All \(\mdot(\jout)\) curves lie on a similar asymptote for small \(\jout\), regardless of \(\rB\).
When \(\jout\) is small enough (\(\jout \lesssim 10\)), the mass accretion rate \(\mdot \propto \jout^{-0.9}\), regardless of \(\rB\). The grey solid straight line in Figure \ref{fig:mdotrb}b represents \( \mdot = 1.6\jout^{-0.9}\). 
Accretion flow with higher angular momentum (\(\jout > 10\)) show dependence on \(\rB\): flows with larger \(\rB\) have higher mass accretion rates for a given \(\jout\). 
The amount of physical angular momentum to be removed is roughly \(\jout\). When \(\rB\) is larger, the flow has enough time (or range in \(r\)) during infall to remove angular momentum, while \(\rB\) is smaller, the flow does not have that time. Therefore, for the same \(\jout\) flow, the mass accretion rate is higher for larger \(\rB\). So again, high angular momentum flows are better represented by \(\lambda\), while low angular momentum flows are better represented by \(\jout\).

Finally, we suggest an approximate functional form for the mass accretion rate in terms of angular momentum, viscosity, and Bondi radius. Figure \ref{fig:mdotrb4} shows the mass accretion rate as a function of the angular momentum \(\jout\) for accretion flows with various viscosity parameters (\(\alpha = 0.1 \sim 0.003\)) at three Bondi radii. The different line styles denote different Bondi radii: dotted lines for \(\rB = 1.0 \times 10^5 \rs\), dashed lines for \(\rB = 1.0 \times 10^4 \rs\), and long-dashed lines for \(\rB = 2.5 \times 10^3 \rs\).

When the angular momentum \(\lambda\) is in the range of \(0.1\) to \(1\), corresponding to high angular momentum, \(\mdot\) can generally be represented by Equation (\ref{eq:P09}):
\begin{equation}
    \mdot \simeq 9 \alpha \lambda^{-p} \;\;\; \mathrm{for } \; 0.1 \lesssim \lambda \lesssim 10,
\label{eq:gen_P09}
\end{equation}
where \(p\) is \(0.40, 0.55, 0.70\), and \(0.84\) for \(\rB = 1.0 \times 10^5, 1.0 \times 10^4, 2.5 \times 10^3\), and \(1.0 \times 10^3 \rs\), respectively.
On the other hand, when the angular momentum \(\jout\) is small enough (\(\jout \lesssim 10\)), the mass accretion rate is approximated by
\begin{equation}
    \mdot \propto \jout^{-p} \;\;\; \mathrm{for } \; \jout \lesssim 10,
\label{eq:eq_mdot_jout}
\end{equation}
where \(p\) is \(0.48, 0.90, 1.31\), and \(1.88\) for \(\alpha = 0.1, 0.03, 0.01\), and \(0.003\), respectively. 

\section{Accretion with Outflow}
\label{sec:ADIOS}

\subsection{Flow Properties}
\begin{figure}
    \centering
    \includegraphics[width=0.8\columnwidth]{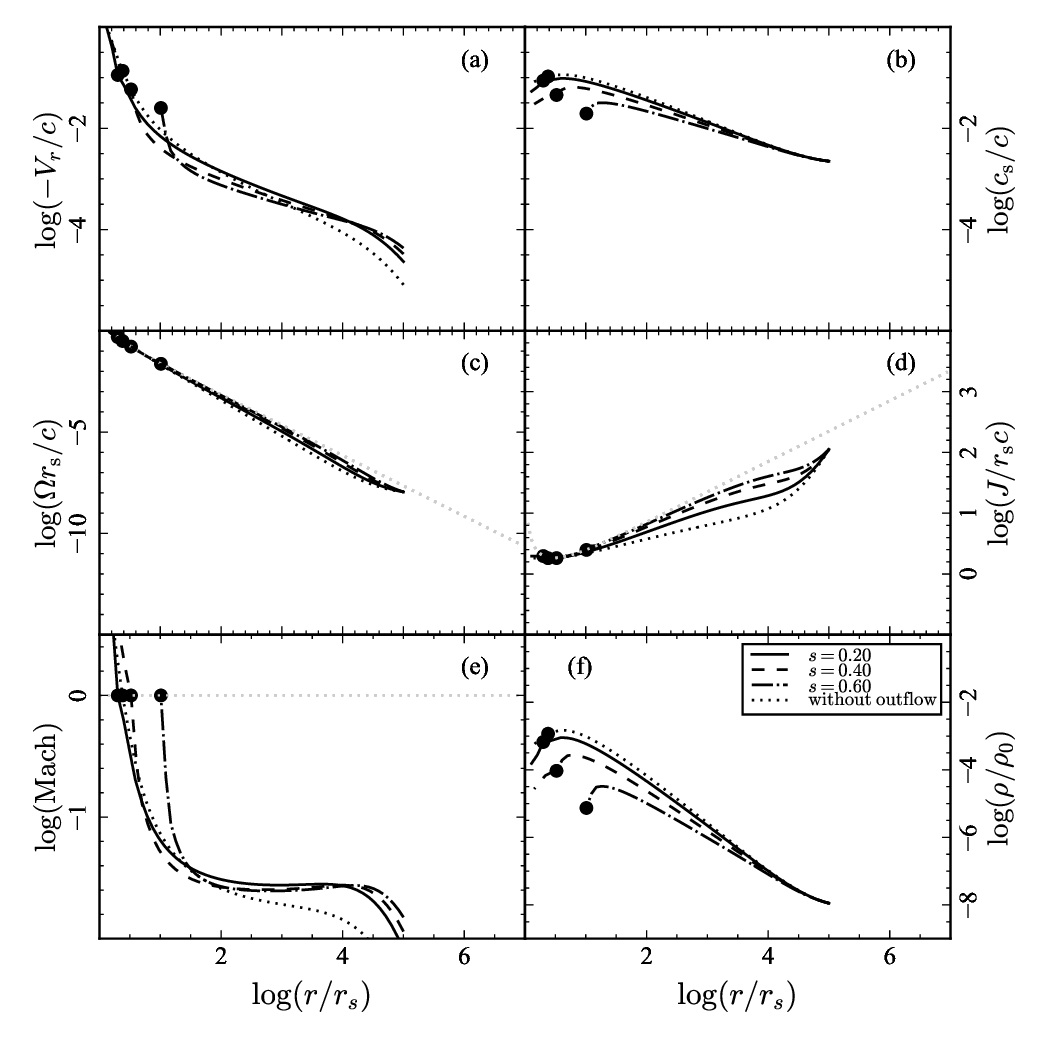}
    \caption{Profiles of accretion flows with angular momentum \(\lambda = 0.5\) and \(\alpha = 0.01\) at \(1.0 \times 10^5 \rs\). Solid, dashed, and dot-dashed lines correspond to ADIOS flows with \(\lBB = 1/2\) for \(s = 0.20\), \(0.40\), and \(0.60\), respectively. The dotted line shows the generalized Bondi flow (without outflow). The filled circles denote the positions of the critical points.}
    \label{fig:profile_adios}
\end{figure}
\begin{figure}
    \centering
    \includegraphics[width=0.9\columnwidth]{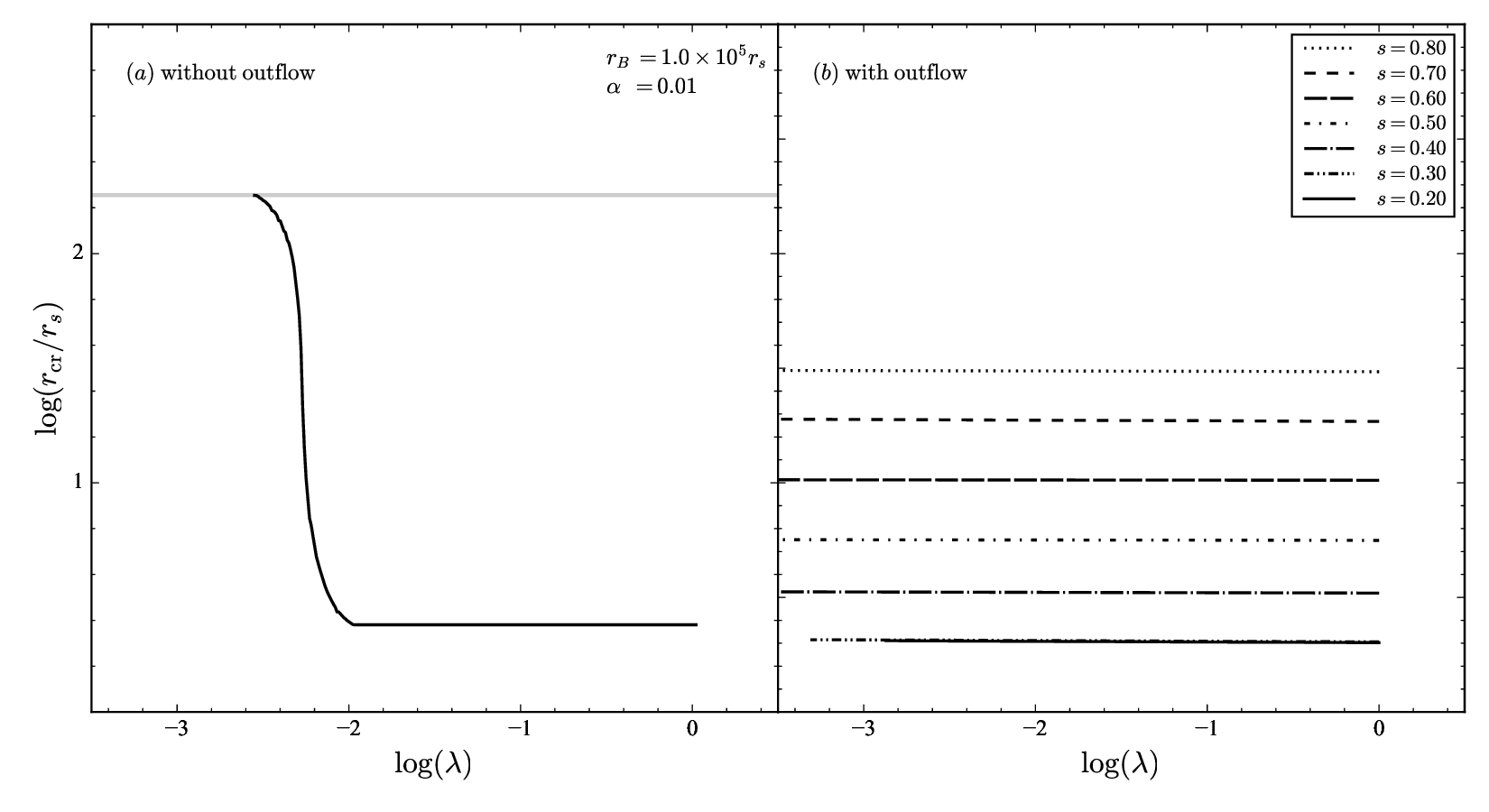}
    \caption{The critical points \(\rcr\) as a function of angular momentum \(\lambda\) for accretion flows with \(\alpha = 0.01\) and \(\rB = 10^5 \rs\). The left and right panels show solutions without and with outflows, respectively. The solutions in the right panel have an outflow parameter \(\lBB = 1/2\) and outflow radial indices s of 0.2, 0.3, 0.4, 0.5, 0.6, 0.7, and 0.8.}
    \label{fig:rcr_adios}
\end{figure}

As is well-known, radiatively inefficient hot accretion flows such as ADAFs have a positive Bernoulli constant \citep{NY1994,NY1995}, which can easily develop winds or outflows that carry away mass, angular momentum, and energy. 
We implement the self-similar ADIOS of BB99 to the generalized Bondi flow to explore its behavior and the mass accretion rate in the presence of outflows. In ADIOS, the mass inflow rate satisfies 
\begin{equation}
    \mdot (r) \propto r^s
 \label{eq:adios_rs}
\end{equation} 
with \(s \geq 0\), where the outflow radial index \(s\) represents the strength of the outflow. Hence, the mass accretion rate at \(r_\mathrm{ISCO}\) is given by \(\mdot(r_\mathrm{ISCO}) = \mdoto(r_\mathrm{ISCO} / \rout)^s\), where \(\mdoto \equiv \mdot(\rout)\).
Recent elaborate multi-dimensional MHD simulations of accretion flows inflowing from well outside the Bondi radius down to the Schwarzschild radius show that the time- and angle-averaged mass accretion rate is well approximated by \(\mdot \propto r^{1/2}\), i.e., \(s = 1/2\) \citep{Guo2024}. Flows with different conditions and physical processes may well be described by different values of \(s\).

The equations of mass and angular momentum conservation in a steady-state ADIOS are 
\begin{equation}
\begin{aligned}
   &\dfrac{\partial \mu v}{\partial r} = 0, \\
   &\dfrac{\partial \mu v J}{\partial r} - \dfrac{\partial G}{\partial r} = 0,
\end{aligned}
 \label{eq:adios_eq}
\end{equation}
where \(\mu\), \(J\), and \(G\) are the mass per unit radius, the specific angular momentum, and the torque exerted on the inner disk by the outer disk, respectively (BB99).
We set the non-Keplerian rotation and torque-zero condition at the Schwarzschild radius \(\rs\). The angular momentum equation is given by
\begin{equation}
     \mu v J - \mu_0 v_0 \JO - G = F_J
\label{eq:adios_eq2}
\end{equation}
where \(F_J\) is the inwardly directed angular momentum flux. The torque \(G\) is described by our choice of viscosity prescription: \(G = \alpha \mu r^3 c_s (d\Omega/dr)\). If the inward flux of angular momentum satisfies \(F_J = \lBB \mdot J\), the angular momentum equation becomes
\begin{equation}
    \dfrac{d\Omega}{dr} = \dfrac{v_r}{\alpha r^3 \cs} \left\{(1-\lBB)\Omega r^2 - \left(\dfrac{\rs}{r}\right)^s \JO\right\},
\label{eq:adios_ang}
\end{equation}
where \(\lBB\) is the parameter that quantifies the loss of angular momentum by the outflow.

We found global solutions for Equations (\ref{eq:adios_eq}) and (\ref{eq:adios_ang}) using the same relaxation method. For convenience, we call these solutions ADIOS flows and the solutions without outflow as explored in \(\S 3\) ADAF. Figure \ref{fig:profile_adios} shows the profiles of radial velocity, sound speed, angular velocity, angular momentum, mach number, and density for ADIOS flows with \(\gamma = 5/3\), \(\alpha = 0.01\), \(\lambda = 0.5\), \(\rB = 1.0 \times 10^5 \rs\), and \(\lBB = 1/2\). Different line styles denote ADIOS flows with different outflow radial indices \(s\): solid lines for \(s=0.2\), dashed lines for \(s=0.4\), and dot-dashed lines for \(s=0.6\). Dotted lines show the generalized Bondi flow (without outflow). Filled circles mark the positions of the critical points. The behaviors of the ADIOS flows resemble those of disk-like flows, and the flows pass a similar critical point close to the inner boundary, the ISCO.

\begin{figure}
    \centering
    \includegraphics[width=0.35\textwidth]{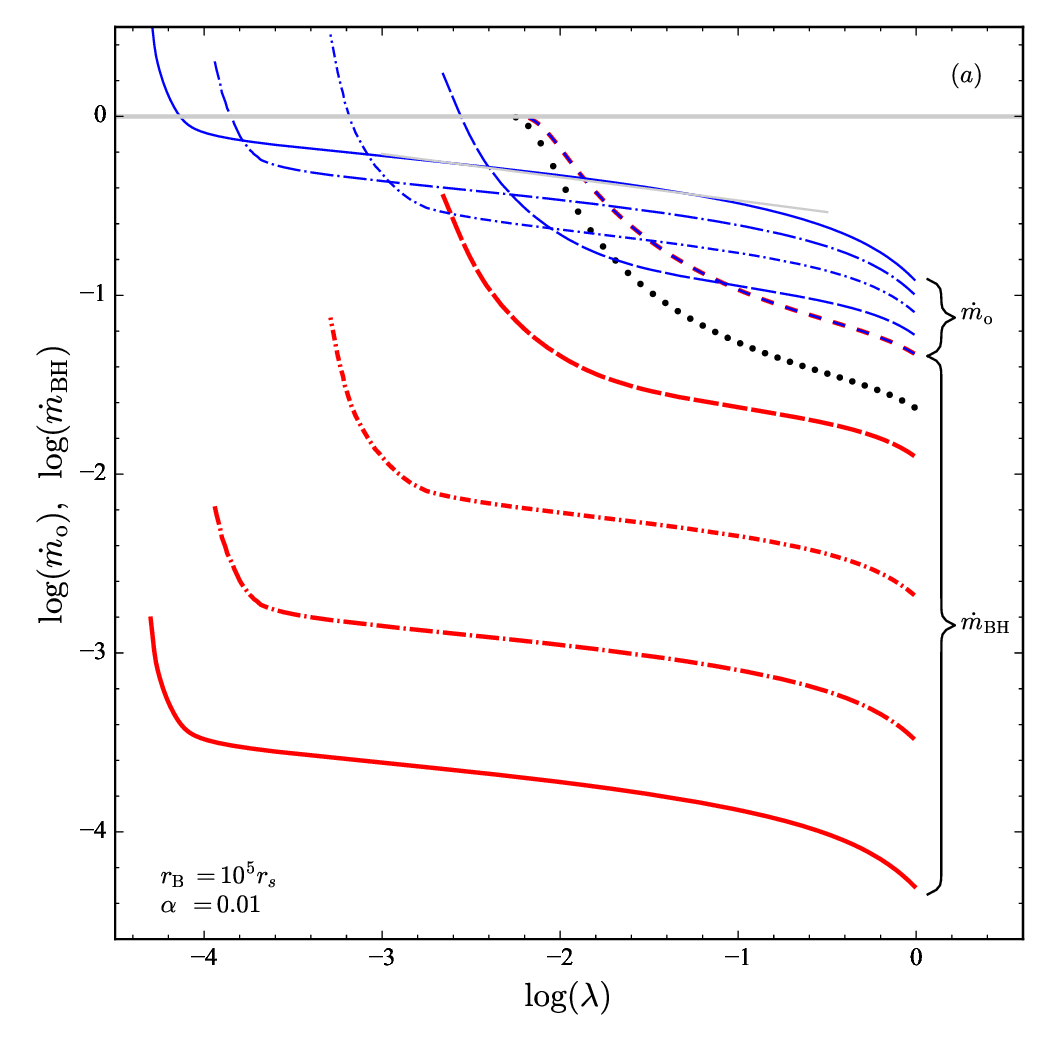}\\
    \includegraphics[width=0.35\textwidth]{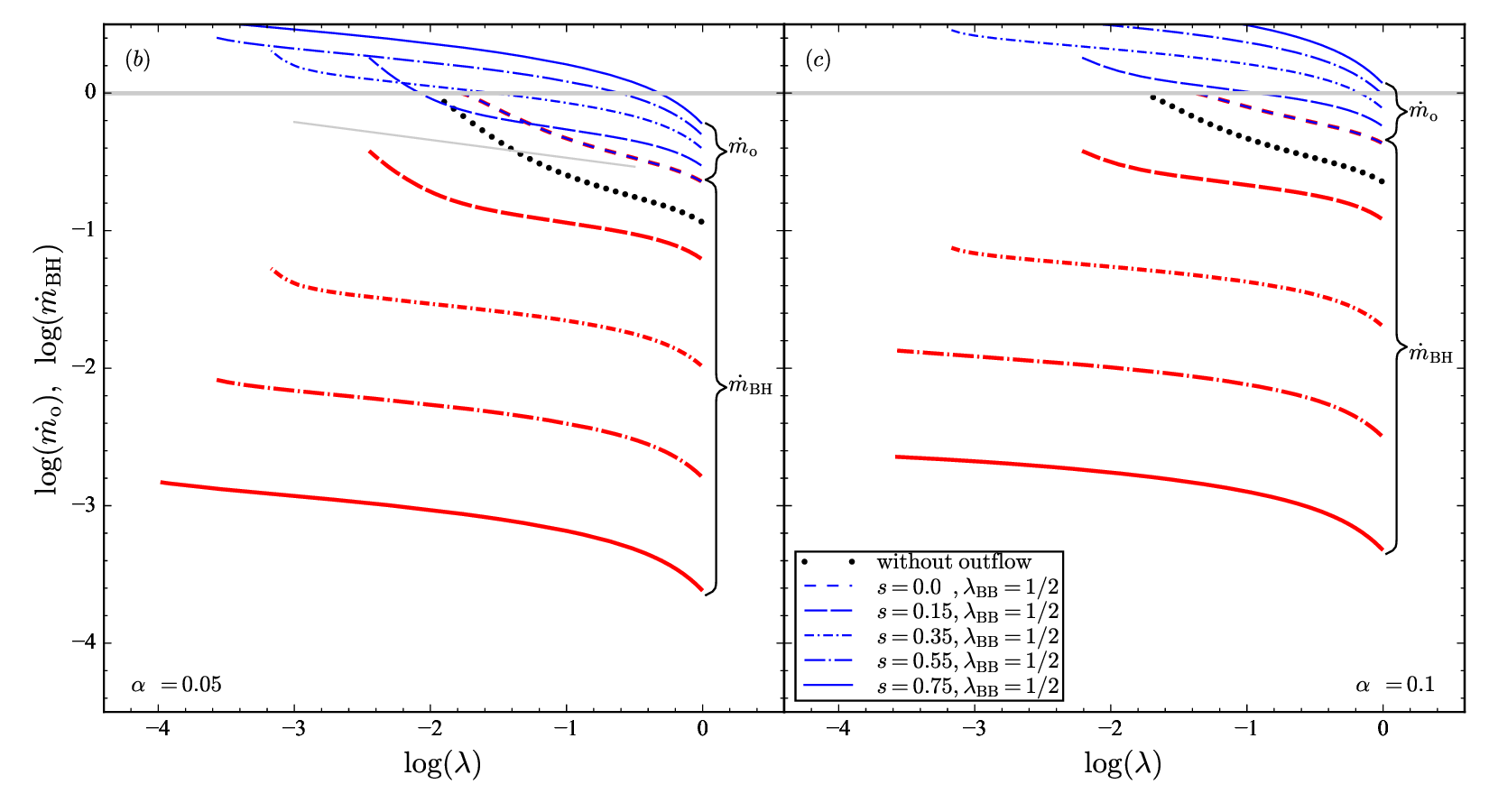}
    \caption{The mass inflow rate \(\mdoto\) for generalized Bondi flows with and without the outflows, as a function of angular momentum, for different values of \(s\) with \(\lBB=1/2\), all at \(\rB = 10^5 \rs\) and \(\alpha = 0.01\) (panel a), \(0.05\) (panel b), \(0.1\) (panel c). Dotted lines denote generalized Bondi flows, as shown in Figure \ref{fig:dep_AM}. Dashed, long-dashed, dot-dashed, dot-long-dashed, and solid lines denote ADIOS flows with \(s = 0.0\), \(0.15\), \(0.35\), \(0.55\), and \(0.75\), respectively.}
    \label{fig:adios}
\end{figure}

The basic effect of outflow is the decrease of the mass inflow rate as \(r\) decreases. Since the radial velocity is limited by the free-fall velocity, at a given radius the density of ADIOS is lower than that of the rotating Bondi flow without outflow (ADAF, hereafter). As the radial velocity of classic Bondi flow is the result of the radial gravitational acceleration versus the pressure gradient force, the radial velocity in ADAF is the result of balance between the gravity, pressure gradient force, and rotational support. Lower density decreases the pressure gradient force, and the radial velocity in ADIOS is higher than that in ADAF, which is shown in Figure \ref{fig:profile_adios}a. 

The specific angular momentum profile changes as well. The absolute magnitude of the term \(v_r / (\alpha r^3 c_s)\) in Equation (\ref{eq:adios_ang}) increases mainly due to increased \(|v_r|\), and \(\Omega(r)\) increases faster than in ADAF (Figure \ref{fig:profile_adios}c). This makes specific angular momentum \(\Omega r^2\) decrease more slowly than in ADAF, and the effect is greater for stronger outflow, i.e., larger \(s\) (Figure \ref{fig:profile_adios}d). As a result, all ADIOS flows rapidly lose their angular momentum near the outer boundary, but at a slower rate than in ADAFs. In the intermediate region (\(\rcr < r < 0.2\rB\)), the angular momentum curve gradually approaches the Keplerian angular momentum (gray dashed line). An ADIOS flow (solid line) with weak outflows (\(s = 0.20\)) quickly loses \(83\%\) of its angular momentum at \(10^4\rs\). On the other hand, an ADIOS flow with relatively strong outflows (\(s = 0.60\)) gradually loses its angular momentum. Stronger outflows keep flows closer to Keplerian. The angular momentum of ADIOS flows with stronger outflows decreases more slowly than that of flows with weak or without outflows, but the critical point is farther out due to the increased radial velocity.

Figure \ref{fig:rcr_adios} shows the locations of the critical points \(\rcr\) as a function of the angular momentum \(\lambda\), for the accretion flows with \(\alpha=0.01\) and \(\rB=10^5\rs\): Figure \ref{fig:rcr_adios}a for ADAF and Figure \ref{fig:rcr_adios}b for ADIOS with outflow. The critical points for the low angular momentum \((\lambda < 10^{-2.5})\) ADAF are similar to those of the classic Bondi solution (dotted horizontal line in Figure \ref{fig:rcr_adios}a). For \(\lambda > 10^{-2}\), the critical points are located close to the ISCO. As noted by NF11, the solutions show a sudden transition from Bondi-like to disk-like flow. Meanwhile, ADIOS flows do not show such a sudden transition. The angular momentum profiles at small radii where the critical points are formed are quite similar regardless of \(\lambda\) (Figure \ref{fig:profile_adios}d), unlike in ADAF (Figure \ref{fig:profiles}d). Figure \ref{fig:rcr_adios}b shows the locations of the critical point of the ADIOS flows with radial indices in the range \(0.3 \leq s \leq 0.8\). The ADIOS flows with a given outflow radial index have a fixed critical point regardless of the angular momentum. The location of the critical point is rather determined by the outflow radial index \(s\). The critical point moves outward with increasing \(s\), i.e., stronger outflow. 

The change in the flow structure due to the presence of outflow is also confirmed in the flow profiles. Without outflows, the flow profile of the accretion flow with low angular momentum (Figure \ref{fig:profiles}) is similar to that of the Bondi flow, whereas the ADIOS flow in Figure \ref{fig:profile_adios} show only a disk-like flow profile even at low angular momentum.

\subsection{Mass Inflow Rate and Accretion Rate}

In the presence of outflow, mass inflow rate \(\mdot(r)\) is not constant but a function of \(r\). The ADIOS flow has a mass inflow rate \(\mdot(r) \propto r^s\) with positive \(s\): mass inflow rate decreases as \(r\) decreases. Only a small fraction of the mass inflow at the outer boundary, which we denote as \(\mdoto\), will reach the inner boundary of the accretion flow \((r_\mathrm{ISCO})\) and finally be accreted into the black hole, the rate of which we denote \(\mdotBH \equiv \mdot(r_\mathrm{ISCO})\). This rate, \(\mdotBH\), is the actual mass accretion rate added to the black hole. 

Figure \ref{fig:adios} shows the mass inflow rate at the outer boundary \(\mdoto\) and the black hole mass accretion rate \(\mdotBH\) for ADAF and ADIOS flows, both for \(\alpha = 0.01\), as a function of angular momentum \(\lambda\), under the same physical conditions but with different outflow strength. The mass inflow rate \(\mdoto\) without outflow is denoted by the thick dotted line. The short dashed line represents the ADIOS flow in which half of the angular momentum \((\lBB = 1/2)\) is carried away, yet the mass inflow rate is conserved. This corresponds to accretion with a magnetic wind (\(s = 0.0\), \(\lBB = 1/2\)) as studied by BB99. The mass inflow rate \(\mdoto\) of the ADIOS flows with high angular momentum is roughly a factor of two higher than that without outflow. These flows are essentially the same as ADAFs with half (or \(\lBB\) times) the value of angular momentum.

\begin{figure}
    \centering
    \includegraphics[width=0.8\columnwidth]{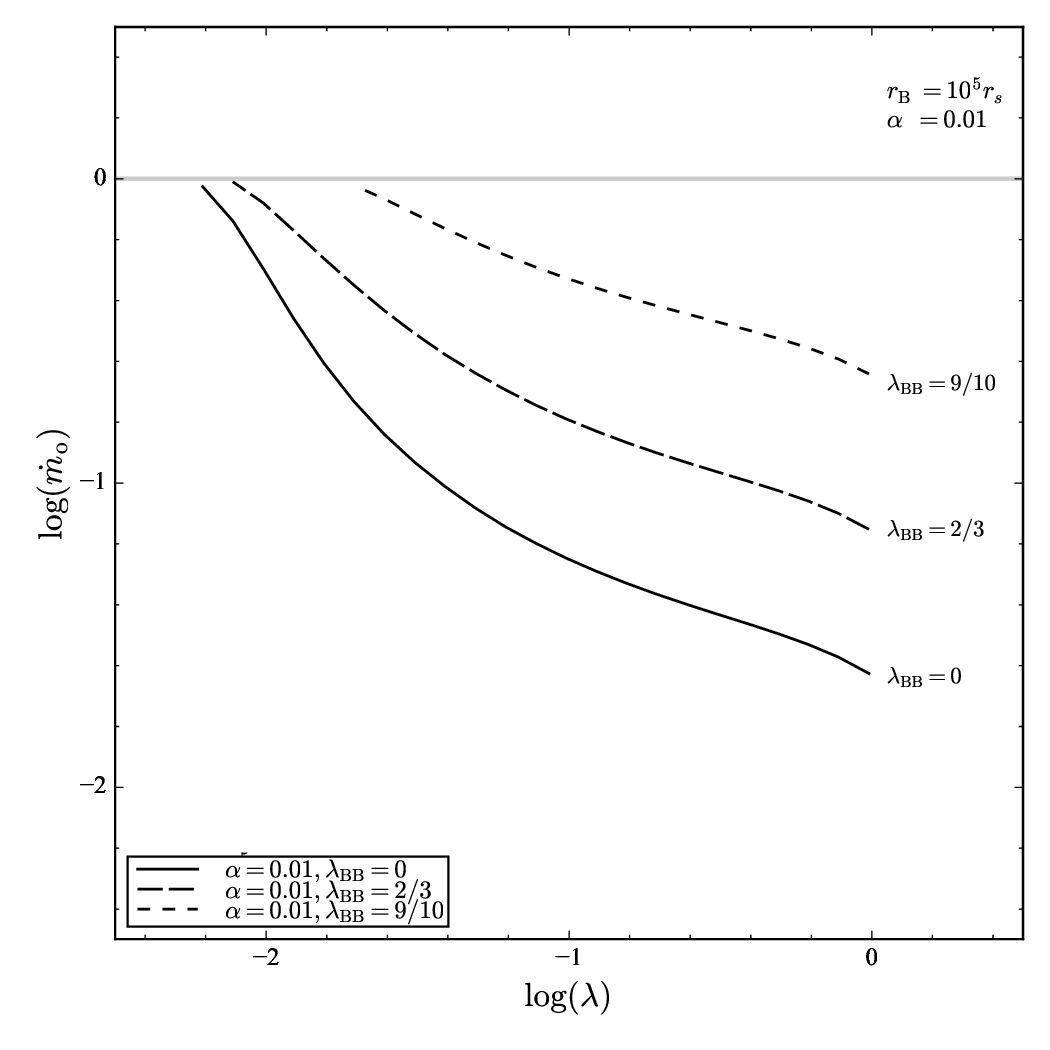}
    \caption{The mass inflow rate at the outer boundary \(\mdoto\) of ADIOS (\(\rB = 1.0 \times 10^5 \rs\), \(\alpha = 0.01\)) as a function of angular momentum \(\lambda\) with different values of \(\lBB\), \(\lBB = 0\), \(2/3\), and \(9/10\) from below.}
    \label{fig:mdot_am_LBB}
\end{figure}

Other lines denote accretion flows with different \(s\): long-dashed line for \(s=0.15\), dot-dashed line for \(s=0.35\), dot-long-dashed line for \(s=0.55\), and solid line for \(s=0.75\). Larger \(s\) means a smaller inflow rate at small radii, i.e., stronger mass outflow and hence a smaller \(\mdotBH\).
The solutions show the effects of outflow on the mass inflow rate at the outer boundary. 
The effect is nontrivial. The outflow decreases the amount of mass whose angular momentum should be removed via viscous stress. However, the outflow also carries away angular momentum, thereby affecting the whole flow structure.

The maximum mass inflow rate of ADAF is the classic Bondi rate (P09, NF11). The mass inflow rate \(\mdoto\) of ADIOS flow, however, can be greater than the Bondi rate, as shown in Figure \ref{fig:adios} because the structure of such flow well inside the Bondi radius can still be that of the flow with mass inflow rate much below the Bondi rate. 

Now we discuss the dependence of the mass inflow rate on the gas angular momentum. The slope of \(\log\mdoto\) versus \(\log\lambda\) in ADIOS becomes successively shallower as \(s\) increases, i.e., as outflow strengthens, compared to ADAFs. This means that the transition from disk-like flow to Bondi-like flow occurs at successively smaller values of the angular momentum \(\lambda\) as \(s\) increases (stronger outflow), relative to the accretion flow without outflow. Consequently, the mass inflow rate at the Bondi radius increases relative to that without outflow for large \(\lambda\), but decreases at small \(\lambda\), and approaches the Bondi rate at a much smaller \(\lambda\). Hence, the outflow can either enhance (for large \(\lambda\)) or weaken (for small \(\lambda\)) the mass inflow rate at the Bondi radius.

\begin{figure}
    \centering
    \includegraphics[width=0.8\columnwidth]{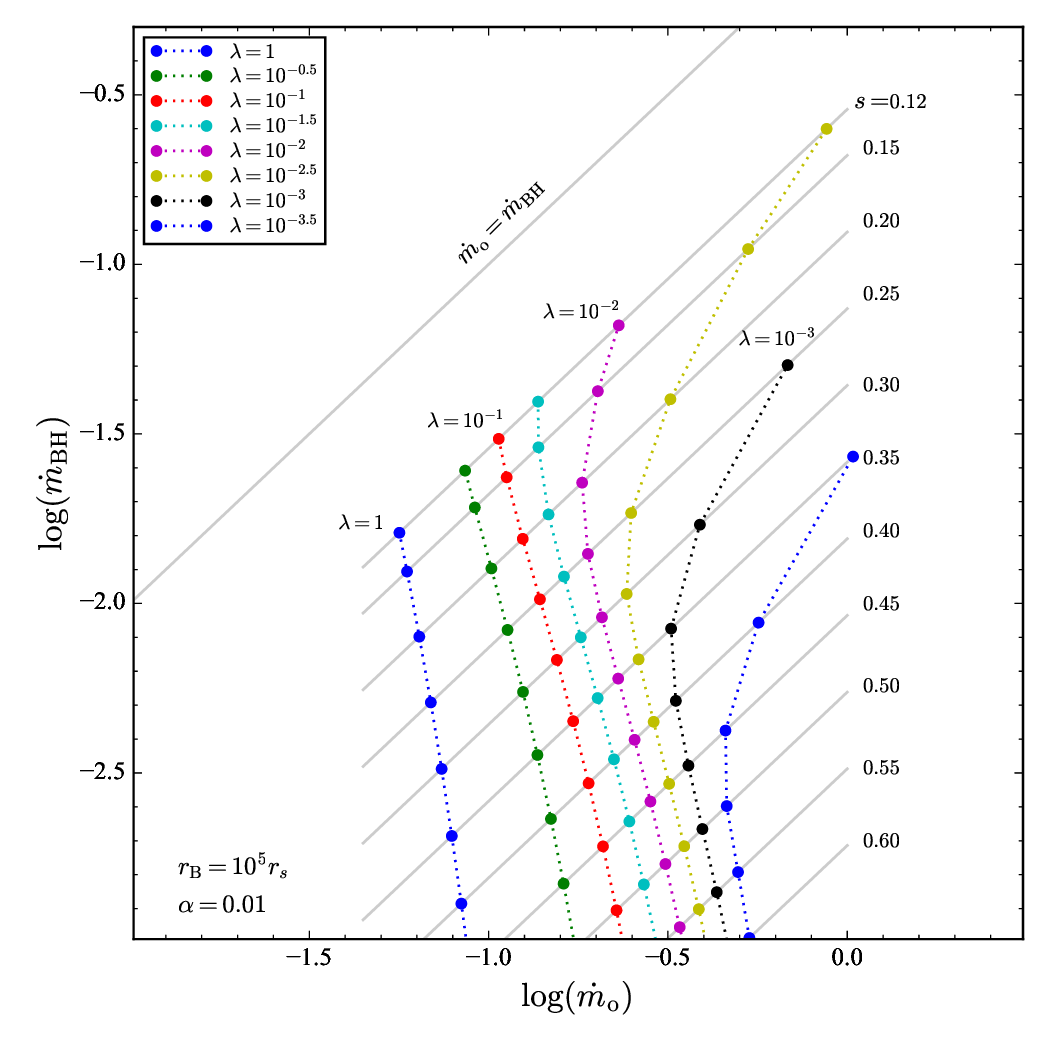}
    \caption{The mass accretion rate \(\mdotBH\) versus the mass inflow rate \(\mdoto\) for ADIOS flows with \(\rB=10^5 \rs\), \(\alpha=0.01\), and \(\lBB=1/2\). Solid lines denote accretion flows with different power-law indices \(s\), while dashed lines denote those with different angular momenta \(\lambda\).}
    \label{fig:mdot_mbh}
\end{figure}

The mass accretion rate obviously depends on the viscosity parameter \( \alpha \) as well. Magnetohydrodynamical simulations suggest \(\alpha \sim 0.01 - 0.003\) \citep{Hawley2011} or \(0.05 - 0.2\) \citep{HB2002}, possibly depending on the magnitude of the net initial magnetic field (see \citealt{YN2014} for further extensive discussion). Figure \ref{fig:adios} shows our global solutions of \(\lBB = 0.5\) for \(\alpha = 0.01\) (panel a), \(0.03\) (panel b), and \(0.1\) (panel c). Higher value of alpha means more efficient removal of angular momentum, and both the mass infall rate at the outer boundary (\(\mdoto\)) and the black hole mass accretion rate (\(\mdotBH\)) increase: the increase of both \(\mdoto\) and \(\mdotBH\) with the increase of the value of s is roughly the same regardless of the value of \(\alpha\). The ADIOS flows with a large value of \( \alpha \) can have a mass infall rate at the outer boundary \(\mdoto\) much greater than the Bondi accretion rate as shown in Figure \ref{fig:adios}b and \ref{fig:adios}c even when the gas angular momentum at the outer boundary is near or almost Keplerian. 

The flow becomes Bondi-like at a higher angular momentum \(\lambda\) reaching the Bondi accretion rate with the increase of \(\alpha\), as expected. However, the slope of \(\log \mdoto\) versus \(\log \lambda\) is less affected by the strength of the outflow. For \(\alpha = 0.03\), the slope becomes slightly shallower, then steeper as \(s\) increases. For \(\alpha = 0.1\), the mass inflow rate is already close to the Bondi accretion rate, and the slope hardly changes as \(s\) increases. So it seems that the strength of the outflow affects low viscosity flows more than high viscosity flows.

Now we explore the effect of angular momentum carried out by the outflow, which is parameterized by \(\lBB\).  If we define \( \alpha^\prime \equiv \alpha/(1-\lBB) \) and \( \JO^\prime \equiv \JO/(1-\lBB) \), Equation (\ref{eq:adios_ang}) becomes 
\begin{equation}
    \dfrac{d\Omega}{dr} = \dfrac{v_r}{\alpha^\prime r^3 \cs} \left\{\Omega r^2 - \left(\dfrac{\rs}{r}\right)^s \JO^\prime\right\}.
    \label{eq:adios_ang2}
\end{equation}

Hence, ADIOS with \(\lBB \neq 0\), \(\alpha\), and \(\JO\) is equivalent to ADIOS with \(\lBB = 0\), \(\alpha^\prime\), and \(\JO^\prime\). Since \(\JO^\prime\) is an eigenvalue determined by the boundary and regularity conditions, it is just a relabeling of \(\JO\). Figure \ref{fig:mdot_am_LBB} is redrawn from Figure \ref{fig:dep_AM} to show how \(\lBB\) changes the mass inflow rate \(\mdoto\): top, middle, and bottom lines for \(\lBB = 0\), \(2/3\), and \(9/10\), respectively. ADIOS with a larger value of \(\lBB\) is equivalent to ADIOS with a larger viscosity parameter \(\alpha^\prime\), and the mass inflow rate increases for a given angular parameter \(\lambda\) and the same boundary conditions, such as the density and temperature of gas at the Bondi radius. The flow also becomes Bondi-like at a higher value of gas angular momentum compared to ADIOS with zero \(\lBB\).

We now compare how much mass in the presence of the outflow is actually accreted into the black hole versus mass inflow rate at the outer boundary. For the ADIOS flow where \(\mdot(r) \propto r^s\), \(\mdotBH\) is simply \(\left(r_\mathrm{ISCO}/\rB\right)^s \cdot \mdoto\). However, \(\mdoto\) is determined by the global structure of the flow. 
The gray diagonal lines in Figure \ref{fig:mdot_mbh} show this relation for different values of \(s\) ranging from \(0.12\) to \(0.60\). For positive \(s\) and \(\rB \gg r_\mathrm{ISCO}\), \(\mdotBH\) can be many orders of magnitude smaller than \(\mdoto\). 
The points in Figure \ref{fig:mdot_mbh} show \(\mdotBH\) as a function of \(\lambda\) for our ADIOS flow solutions with outflow radial indices \(0.12 \leq s \leq 0.60\) and \(\rB=1.0\times10^5\rs\). For example, with \(\rB = 10^5 \rs\), \(\lBB = 1/2\), and \(s = 0.6\), \(\mdoto = 0.1\) for a \(\lambda = 1\) (nearly Keplerian) flow, but \(\mdotBH = 10^{-2.7} \mdoto\). Hence, only \(\mdotBH \simeq 1.9 \times 10^{-4}\) contributes to the growth of the black hole.
However, for a weaker outflow with \(s = 0.12\) (and \(\lambda = 1\)), \(\mdoto \simeq 10^{-1.25}\) and \(\mdotBH \simeq 10^{-1.8}\): the ratio \(\mdotBH/\mdoto\) is modest. ADIOS flows with relatively high angular momentum (\(\lambda \gtrsim 0.1\)) show similar behavior. Stronger outflows (larger \(s\)) modestly increase \(\mdoto\) compared to no-outflow solutions but greatly decreases \(\mdotBH\).
ADIOS flows with lower angular momentum (\(\lambda \lesssim 0.1\)) show similar properties until they become Bondi-like. As shown in Figure \ref{fig:adios}, stronger outflows cause ADIOS flows to become Bondi-like at much smaller values of the angular momentum \(\lambda\). Therefore, for the same \(\lambda = 10^{-2.5}\) (light green points), an ADIOS flow with a weaker outflow, such as \(s = 0.12\), is already Bondi-like with \(\mdoto \sim 1\), whereas an ADIOS flow with a stronger outflow, such as \(s = 0.55\), is not yet Bondi-like and has \(\mdoto \simeq 10^{-0.42}\). 
Hence, for ADIOS flows with low angular momentum (\(\lambda \ll 1\)), stronger outflows first decrease and then increase \(\mdoto\), but always decrease \(\mdotBH\) by many orders of magnitude. 

\section{Summary \& Discussion}
\label{sec:summary}

We have constructed global solutions for the generalized Bondi flow (i.e., rotating polytropic accretion flow) under various physical conditions and explored the dependence of flow characteristics on the viscosity parameter, gas angular momentum, and Bondi radius (or equivalently, the gas temperature at the Bondi radius). When the mass accretion rate is expressed in units of the classic Bondi accretion rate, the Bondi radius in units of the Schwarzschild radius, and the gas angular momentum in units of the Keplerian angular momentum (at the Bondi radius), our solutions show that the mass accretion rate always decreases with increasing gas angular momentum and with increasing Bondi radius. For a small Bondi radius (i.e., high gas temperature), the mass accretion rate is roughly inversely proportional to the angular momentum and proportional to the viscosity parameter \(\alpha\) for not too small angular momentum, as shown in P09. However, for accretion flows with a larger Bondi radius (i.e., lower gas temperature at the Bondi radius), the decrease in the mass accretion rate with increasing angular momentum is successively weaker, and the flow becomes Bondi-like only at a much smaller angular momentum. Therefore, the decrease in the mass accretion rate relative to the Bondi accretion rate is modest when the Bondi radius is large, which is noted by NF11. In general, the decrease in the mass accretion rate depends not only on the gas angular momentum but also on the Bondi radius. We additionally find that the dependence of the mass accretion rate on the gas angular momentum becomes increasingly independent of the Bondi radius when the gas angular momentum expressed in units of \(\rsc\) is \(\lesssim 10\), and that the critical gas angular momentum below which the mass accretion rate approaches the Bondi accretion rate is determined (in units of \(\rsc\)) only by the viscosity parameter \(\alpha\). 

We have also studied the generalized Bondi flow that includes outflows, as described by the ADIOS model. For a given radial dependence of the outflow, this ADIOS flow shows disk-like characteristics regardless of the angular momentum at the outer boundary. Our global solutions show that ADIOS flows have up to five times higher mass inflow rates at the outer boundary compared to those without outflows, for the same gas density, temperature, and angular momentum at the outer boundary. The radial velocity of ADIOS flows increases with the strength of the outflow. Strong outflow decreases the gas density as radius decreases. This causes the pressure gradient force weaker, and the radial velocity increases. This results in a significant increase in the mass inflow rate at the outer boundary, which can be greater than the Bondi accretion rate. Stronger outflow also weakens the dependence of the mass accretion rate on the gas angular momentum. The removal of angular momentum by outflow works as if the viscosity has increased, yielding a higher mass inflow rate at the outer boundary. 

Despite the increase in the mass inflow rate at the outer boundary, however, the rate of the accreted mass that actually adds to the black hole mass is many orders of magnitude smaller than the mass inflow rate at the outer boundary, depending on the strength of the outflow. The amount of energy production by ADIOS flow will be determined by the mass inflow rate at the outer boundary, the final mass accretion rate to the black hole, and all the detailed physical and radiative processes that occur within the accretion flow. Hence, the energy production from an accreting black hole will depend on the gas angular momentum at the outer boundary, the presence of outflows, the strength of the magnetic field that may determine the viscosity as well as the density and temperature of gas at the outer boundary as in classical Bondi flow. 
We want to add that our \textquoteleft global \textquoteright ADIOS flow solutions bear the assumption of a self-similar model of the outflow, and that the detailed physical characteristics of the flow may differ from those in more realistic multi-dimensional flows where outflows are naturally produced from basic physical processes.
Yet, our result implies that when a black hole is accreting gas with given density and temperature, the mass inflow rate from the surrounding gas and the mass accretion rate added to the black hole mass can be widely different from the classic Bondi accretion rate, depending on the amount of gas angular momentum and the strength of the outflow. Therefore, the growth of the black hole and the energy production from the accretion will also be very different from those based on the classic Bondi accretion result. 

\section*{Acknowledgements}
We thank the anonymous reviewer for many insightful comments which have greatly improved our paper. This research was supported by the National Research Foundation of Korea (NRF) funded by the Ministry of Education (No-RS-2019-NR045193, RS-2018-NR031074). DHH acknowledges the support from the National Research Foundation of Korea (No-2022R1A4A3031306).


\begin{thebibliography}{}
\bibitem[Abramowicz et al.(1988)]{Ab1988} Abramowicz, M.~A., Czerny, B., Lasota, J.~P., \& Szuszkiewicz, E.\ 1988, \apj, 332, 646. doi:10.1086/166683
\bibitem[Baganoff et al.(2003)]{Baganoff2003} Baganoff, F.~K., Maeda, Y., Morris, M., et al.\ 2003, \apj, 591, 891. doi:10.1086/375145
\bibitem[Begelman(2012)]{Begelman2012} Begelman, M.~C.\ 2012, \mnras, 420, 2912. doi:10.1111/j.1365-2966.2011.20071.x
\bibitem[Blandford \& Begelman(1999)]{BB1999} Blandford, R.~D., \& Begelman, M.~C.\ 1999, \mnras, 303, L1. doi:10.1046/j.1365-8711.1999.02358.x
\bibitem[Bondi(1952)]{Bondi1952} Bondi, H.\ 1952, \mnras, 112, 195. doi:10.1093/mnras/112.2.195
\bibitem[Bower et al.(2018)]{Bower2018} Bower, G.~C., Broderick, A., Dexter, J., et al.\ 2018, \apj, 868, 101. doi:10.3847/1538-4357/aae983
\bibitem[Bu \& Yang(2019)]{BY2019} Bu, D.-F. \& Yang, X.-H.\ 2019, \mnras, 484, 1724. doi:10.1093/mnras/stz050
\bibitem[Event Horizon Telescope Collaboration et al.(2021)]{EHT2021} Event Horizon Telescope Collaboration, Akiyama, K., Algaba, J.~C., et al.\ 2021, \apjl, 910, L13. doi:10.3847/2041-8213/abe4de
\bibitem[Guo et al.(2023)]{Guo2023} Guo, M., Stone, J.~M., Kim, C.-G., et al.\ 2023, \apj, 946, 26. doi:10.3847/1538-4357/acb81e
\bibitem[Guo et al.(2024)]{Guo2024} Guo, M., Stone, J.~M., Quataert, E., et al.\ 2024, \apj, 973, 141. doi:10.3847/1538-4357/ad5fe7
\bibitem[Hawley(2011)]{Hawley2011} Hawley, J.~F.\ 2011, 5th International Conference of Numerical Modeling of Space Plasma Flows (ASTRONUM 2010), 444, 63. 
\bibitem[Hawley \& Balbus(2002)]{HB2002} Hawley, J.~F. \& Balbus, S.~A.\ 2002, \apj, 573, 2, 738. doi:10.1086/340765
\bibitem[Inayoshi et al.(2018)]{Inayoshi2018} Inayoshi, K., Ostriker, J.~P., Haiman, Z., et al.\ 2018, \mnras, 476, 1412. doi:10.1093/mnras/sty276
\bibitem[Kuo et al.(2014)]{Kuo2014} Kuo, C.~Y., Asada, K., Rao, R., et al.\ 2014, \apjl, 783, L33. doi:10.1088/2041-8205/783/2/L33
\bibitem[Li et al.(2013)]{Li2013} Li, J., Ostriker, J., \& Sunyaev, R.\ 2013, \apj, 767, 105. doi:10.1088/0004-637X/767/2/105
\bibitem[Lynden-Bell \& Pringle(1974)]{LP1974} Lynden-Bell, D. \& Pringle, J.~E.\ 1974, \mnras, 168, 603. doi:10.1093/mnras/168.3.603
\bibitem[Marrone et al.(2007)]{Marrone2007} Marrone, D.~P., Moran, J.~M., Zhao, J.-H., et al.\ 2007, \apjl, 654, L57. doi:10.1086/510850
\bibitem[Nakamura et al.(1997)]{Nakamura1997} Nakamura, K.~E., Kusunose, M., Matsumoto, R., \& Kato, S.\ 1997, \pasj, 49, 503. doi:10.1093/pasj/49.4.503
\bibitem[Narayan et al.(2012)]{Narayan2012} Narayan, R., S{\"A} dowski, A., Penna, R.~F., et al.\ 2012, \mnras, 426, 3241. doi:10.1111/j.1365-2966.2012.22002.x
\bibitem[Narayan \& Fabian(2011)]{NF2011} Narayan, R., \& Fabian, A.~C.\ 2011, \mnras, 415, 3721. doi:10.1111/j.1365-2966.2011.18987.x
\bibitem[Narayan \& Yi(1994)]{NY1994} Narayan, R., \& Yi, I.\ 1994, \apjl, 428, L13. doi:10.1086/187381
\bibitem[Narayan \& Yi(1995)]{NY1995} Narayan, R. \& Yi, I.\ 1995, \apj, 444, 231. doi:10.1086/175599
\bibitem[Narayan et al.(1997)]{Narayan1997} Narayan, R., Kato, S., \& Honma, F.\ 1997, \apj, 476, 49. doi:10.1086/303591
\bibitem[Paczy{\'n}sky \& Wiita(1980)]{PW1980} Paczy{\'n}sky, B., \& Wiita, P.~J.\ 1980, \aap, 88, 23
\bibitem[Park(1990a)]{Park1990a} Park, M.-G.\ 1990, \apj, 354, 83. doi:10.1086/168669
\bibitem[Park(1990b)]{Park1990b} Park, M.-G.\ 1990, \apj, 354, 64. doi:10.1086/168668
\bibitem[Park(2009)]{Park2009} Park, M.-G.\ 2009, \apj, 706, 637. doi:10.1088/0004-637X/706/1/637
\bibitem[Park \& Han(2018)]{PH2018} Park, M.-G., \& Han, D.-H.\ 2018, EPJWC, 04005. doi:10.1051/epjconf/201816804005
\bibitem[Park \& Ostriker(2001)]{PO2001} Park, M.-G. \& Ostriker, J.~P.\ 2001, \apj, 549, 100. doi:10.1086/319042
\bibitem[Park \& Ostriker(1999)]{PO1999} Park, M.-G. \& Ostriker, J.~P.\ 1999, \apj, 527, 247. doi:10.1086/308061
\bibitem[Ryu et al.(1997)]{RCM97} Ryu, D., Chakrabarti, S.~K., \& Molteni, D.\ 1997, \apj, 474, 378. doi:10.1086/303461
\bibitem[Shakura \& Sunyaev(1973)]{SS1973} Shakura, N.~I., \& Sunyaev, R.~A.\ 1973, \aap, 24, 337 
\bibitem[Shapiro(1973)]{Shapiro1973} Shapiro, S.~L.\ 1973, \apj, 180, 531. doi:10.1086/151982
\bibitem[Stone et al.(1999)]{Stone1999} Stone, J.~M., Pringle, J.~E., \& Begelman, M.~C.\ 1999, \mnras, 310, 1002. doi:10.1046/j.1365-8711.1999.03024.x
\bibitem[Stone \& Pringle(2001)]{SP2001} Stone, J.~M. \& Pringle, J.~E.\ 2001, \mnras, 322, 461. doi:10.1046/j.1365-8711.2001.04138.x
\bibitem[Turolla \& Dullemond(2000)]{TD2000} Turolla, R. \& Dullemond, C.~P.\ 2000, \apjl, 531, L49. doi:10.1086/312527
\bibitem[Wang et al.(2013)]{Wang2013} Wang, Q.~D., Nowak, M.~A., Markoff, S.~B., et al.\ 2013, Science, 341, 981. doi:10.1126/science.1240755
\bibitem[Xu \& Chen(1997)]{XC1997} Xu, G. \& Chen, X.\ 1997, \apjl, 489, L29. doi:10.1086/310956
\bibitem[Yoon et al.(2020)]{Yoon2020} Yoon, D., Chatterjee, K., Markoff, S.~B., et al.\ 2020, \mnras, 499, 3178. doi:10.1093/mnras/staa3031
\bibitem[Yuan(1999)]{Yuan1999} Yuan, F.\ 1999, \apjl, 521, L55. doi:10.1086/312173
\bibitem[Yuan \& Narayan(2014)]{YN2014} Yuan, F. \& Narayan, R.\ 2014, \araa, 52, 529. doi:10.1146/annurev-astro-082812-141003
\bibitem[Yuan et al.(2000)]{Yuan2000} Yuan, F., Peng, Q., Lu, J.-f., \& Wang, J.\ 2000, \apj, 537, 236. doi:10.1086/309020 
\bibitem[Yuan et al.(2012a)]{Yuan2012a} Yuan, F., Wu, M., \& Bu, D.\ 2012, \apj, 761, 129. doi:10.1088/0004-637X/761/2/129
\bibitem[Yuan et al.(2012b)]{Yuan2012b} Yuan, F., Bu, D., \& Wu, M.\ 2012, \apj, 761, 130. doi:10.1088/0004-637X/761/2/130
\bibitem[Yuan et al.(2015)]{Yuan2015} Yuan, F., Gan, Z., Narayan, R., et al.\ 2015, \apj, 804, 101. doi:10.1088/0004-637X/804/2/101

\end{thebibliography}
\end{document}